\newcommand{\pointsize}{11pt}

\documentclass[oneside, \pointsize]{amsbook}

\usepackage[
   includehead,
   includefoot,
     left = 1.5in, 
      top = 0.5in, 
    right = 1in,
   bottom = 1in
]{geometry}
\usepackage{fancyhdr}
\usepackage{setspace}
\usepackage{calc}

\usepackage{amssymb}
\usepackage{amsthm}
\usepackage{amsmath}
\usepackage{mathrsfs}
\fancyheadoffset[R]{0.5in} 

\fancypagestyle{prelim}{%
   \renewcommand{\headrulewidth}{0pt} 
   \fancyhf{}           
   \pagenumbering{roman}    
   \cfoot{-\thepage-}       
}

\fancypagestyle{maintext}{%
   \renewcommand{\headrulewidth}{0.4pt}
   \pagenumbering{arabic}
   \fancyhf{}
   \fancyhead[L]{\rightmark}
   \rhead{\thepage}
}

\numberwithin{figure}{chapter} 
\numberwithin{table}{chapter}
\numberwithin{equation}{chapter}
\numberwithin{section}{chapter}

\usepackage{graphicx}
\usepackage{amssymb}
\usepackage{amsthm}
\usepackage{amsmath}
\usepackage{mathrsfs}

\newtheorem{thm}{Theorem}[section]

\newtheorem{lem}{\textit{Lemma}}[section]
\newtheorem{cor}{\textit{Corollary}}[section]
\theoremstyle{cor}\newtheorem*{rem}{Remark}
\theoremstyle{plain}
\theoremstyle{definition}\newtheorem{Def}{Definition}[section]
\newcommand{\defeq}{\stackrel{\text{\tiny def}}{=}}

\begin{document}
   \frontmatter

   \pagestyle{prelim}
   
   %
   \fancypagestyle{plain}{%
      \fancyhf{}
      \cfoot{-\thepage-}
   }%
   \begin{center}
   \null\vfill
   \textbf{%
     Codes in $W^*$-metric Spaces: Theory and Examples
   }%
   \\
   \bigskip
   By \\
   \bigskip
   CHRISTOPHER J. BUMGARDNER \\
   \bigskip
   B.S. (Binghamton Univerisity) 2003 \\
   \bigskip
   DISSERTATION \\
   \bigskip
   Submitted in partial satisfaction of the requirements for the
   degree of \\
   \bigskip
   DOCTOR OF PHILOSOPHY \\
   \bigskip
   in \\
   \bigskip
   Mathematics \\
   \bigskip
   in the \\
   \bigskip
   OFFICE OF GRADUATE STUDIES \\
   \bigskip        
   of the \\
   \bigskip
   UNIVERSITY OF CALIFORNIA \\
   \bigskip
   DAVIS \\
   \bigskip
   Approved: \\
   \bigskip
   \bigskip
   \makebox[3in]{\hrulefill} \\
   Greg Kuperberg, Chair \\
   \bigskip
   \bigskip
   \makebox[3in]{\hrulefill} \\
  Bruno Nachtergaele\\
   \bigskip
   \bigskip
   \makebox[3in]{\hrulefill} \\
   Andrew Waldron \\
   \bigskip
   Committee in Charge \\
   \bigskip
   2011 \\
   \vfill
\end{center}

   \newpage
   
   %
   \doublespacing
   
   \tableofcontents
   \newpage
   
   {\singlespacing
   \begin{flushright}
      Christopher J. Bumgardner \\
      March 2011\\
      Mathematics \\
   \end{flushright}
}

\bigskip

\begin{center}
   Codes in $W^*$-metric Spaces: Theory and Examples\\
\end{center}

\section*{Abstract}

\indent\indent We introduce a $W^*$-metric space, which is a particular approach to non-commutative metric spaces where a \textit{quantum metric} is defined on a von Neumann algebra.  We generalize the notion of a quantum code and quantum error correction to the setting of finite dimensional $W^*$-metric spaces, which includes codes and error correction for classical finite metric spaces.  We also introduce a class of $W^*$-metric spaces that come from representations of semi-simple Lie algebras $\mathfrak{g}$ called \textit{$\mathfrak{g}$-metric} spaces, and present an outline for code constructions.  In turn, we produce specific code constructions for $\mathfrak{su}(2,\mathbb{C})$-metric spaces that depend upon proving Tverberg's theorem for points on a moment curve constructed from arithmetic sequences.  We introduce a \textit{quantum distance distribution}, and we prove an analogue of the MacWilliam's identities for $\mathfrak{su}(2)$-metric spaces.

   \newpage
   
   
   \mainmatter
   
   \pagestyle{maintext}
   
   %
   \fancypagestyle{plain}{%
      \renewcommand{\headrulewidth}{0pt}
      \fancyhf{}
      \rhead{\thepage}
   }%
   
   \chapter{Introduction}
   \label{ch:IntroductionLabel}
   \begin{section}{Introduction}
\indent\indent Metric spaces play a fundamental role in classical coding theory.  For example, the set of $n$-tuples of  0's and 1's, called bit strings of length $n$, can be given a metric, the Hamming metric, that counts the number entries that differ in two bit strings (e.g. the distance between $(0,0,1)$ and $(1,1,1)$ is $2$).  A \textit{code} in a metric space $M$, is simply a subset $C\subset M$.  A code $C$ has a minimal distance $d$ if the distance between any two elements of $C$ is no less than $d$.  The use of Hamming metric can be seen in the following example.  Suppose we have two people, Alice and Bob, where Bob asked Alice a ``yes" or ``no" question at some point in time.  Then Alice would like to send Bob an answer using bit strings, but the method of transport of the bit strings is known to possibly change one of the bits in the string.  Thus we would like to encode ``yes" and ``no'' answers into bit strings where the minimal distance of the code is 3, so we can distinguish the answers even if an error occurs.  For example, what Alice and Bob could do is decide before hand that the ``yes" answer with be the string $000$ and the ``no" answer will be $111$.  If Alice sends her response and Bob gets the answer, say $001$, he knows the answer was ``yes".  This is because $001$ is a distance $1$ away from $000$ and a distance $2$ away from $111$, and Bob knew the message he received from Alice would be at most a distance $1$ away from $000$ or $111$.  Other examples of the use of metric spaces in classical coding theory can be found in \cite{SPLG}.  Among those examples are the use of lattices in $\mathbb{R}^n$ for signal processing.  If we turn to quantum coding theory, we do not see any metric spaces.  Although, in quantum coding theory the notion of ``distance" is given to operators on a Hilbert space.  The quantum Hamming ``metric" or filtration being the most popular one which we introduce later in the introduction.

\indent It could be said that quantum information theory has given an example of what should be called a non-commutative or quantum metric space; this being, in analogy to classical information theory, a  fundamental object to quantum information theory.  Although, in what context does the previous statement make sense?  The operator algebra route of generalizing any classical notion of, for example, topological or measure spaces is to begin with understanding how properties of the space correspond to properties of a chosen algebra of functions on the space.   For example, we may choose the complex valued, continuous functions on a topological space $X$, $C(X)$, (an example of a $C^*$ algebra), or $L^{\infty}(X)$ for some measure space $X$ (an example of a von Neumann algebra).  Before elaborating on this point, we recall the definitions of a $C^*$ algebra and a von Neumann algebra (also referred to as a $W^*$-algebra) as these both have been used to define a quantum or non-commutative metric space.   

\begin{Def} A $C^*$  \textit{algebra} $X$ is a Banach algebra over the complex numbers with an involution $*:X\rightarrow X$ that is conjugate-linear and satisfies $(xy)^*=y^*x^*$.  Lastly, the norm on $X$ satisfies the $C^*$  \textit{condition}.  This meaning, 
\[ ||xx^*||=||x|| ||x^*||\]
\end{Def}

\begin{Def} \label{vonNeumann} The following are equivalent definitions of a von Neumann algebra.
\begin{enumerate}
\item A \textit{von Neumann algebra} is a $C^*$ algebra that has a predual.  In other words, as a Banach algebra, it is dual to some other Banach algebra.  \
\item A \textit{von Neumann algebra} is a unital, weakly closed, $*$-closed algebra of bounded operators on a Hilbert space.  \
\item A \textit{von Neumann algebra} is the commutant of a $*$-closed subset of bounded operators on a Hilbert space.   
\end{enumerate}
\end{Def}

\indent To continue, locally compact, Hausdorff topological spaces have a contravariantly functorial equivalence to commutative $C^*$ algebras as seen in the Gelfand representation theorem (see for example \cite{OpAlg}).  To give the reader some intuition into this correspondence, we begin with a commutative $C^*$-algebra $\mathcal{C}$.  We construct a set $X=\{x_M;\,M\,a\,maximal\,ideal\,in\,\mathcal{C}\}$.  Then an ideal $I$ in $\mathcal{C}$ can be considered a subset $C_I\subset X$ via $x_M\in C_I$ iff $I\subset M$.  It then can be shown that the collection of subsets $\{C_I;\,I\,an\,ideal\,in\,\mathcal{C}\}$ satisfies all conditions to be considered a collection of closed subsets of $X$.  Then $X$ with this topology is a locally compact, Hausdorff space.  If $\mathcal{C}$ is unital, then $X$ is compact.  The reverse construction begins with a locally compact, Hausdorff space $X$.  Then the set of compactly supported, continuous functions $C_c(X)$ forms a $C^*$-algebra with the $\sup$ norm.  It is then natural to consider non-commutative $C^*$ algebras as non-commutative locally compact, Hausdorff topological spaces.  

\indent The relationship between commutative von Neumann algebras and measure spaces is entirely analogous to the relationship between $C^*$-algebras and locally compact, Hausdorff topological spaces.  Every commutative von Neumann algebra is isomorphic to $L^{\infty}(X,\mu)$ for some measure space $X$ with measure $\mu$.  To see $L^{\infty}(X,\mu)$ as a sub-collection of operators acting on a Hilbert space one can consider $L^{\infty}(X,\mu)$ acting as multiplication operators on $L^{2}(X,\mu)$.  Also, given a measure space $(X,\mu)$, $L^{\infty}(X,\mu)$ is a commutative von Neumann algebra.  One may note that $L^{\infty}(X,\mu)$ is a unital $C^*$-algebra as well, so it must correspond to $C(Y)$ for some compact, Hausdorff space $Y$.  Although this space is usually very disconnected or Stonean.   

\indent One may now ask about metric information in the context of commutative $C^*$-algebras.  Given a (compact) metric space $M$ with metric $m$, can metric information correspond to properties of $C(M)$, the continuous function on $M$, in a way where those properties of $C(M)$ can be used to reconstruct the metric on $M$?  In \cite{Rie04}, a \textit{Lip-norm} is on a $C^*$ algebra is defined, and it is used to define a \textit{compact quantum metric space}.  For technical clarity the Lip-norm on a simpler object than a $C^*$ algebra, namely an \textit{order unit space} (a generalization of the set of positive elements, $\{A^*A\}$, in a $C^*$ algebra).  The definition of a Lip-norm as defining a \textit{quantum} metric space was partially inspired by Kantorovich's observation (\cite{Kant1}, \cite{Kant2}) that the Lipschitz seminorm on $C(M)$ for compact $M$ can be used to recover the metric $m$.  Given a compact metric space $(M,d)$, the Lipschitz semi-norm on $C(M)$ is defined as (it is allowed to take on infinite values):
\[L(f)=\sup \left\{ \frac{|f(x)-f(y)|}{d(x,y)};\, x\neq y\right\} . \]

Then we can recover the metric $d$ from the Lipschitz semi-norm on $C(M)$ by:
\[d(x,y)=\sup \{|f(x)-f(y)|;\, L(f)\leq 1\}.\]
 
 \indent Kantorovich further showed that for a commutative, unital $C^*$ algebra, $\mathcal{C}$, the Lipschitz semi-norm can be can be used to define a distance on the \textit{normalized state space} of $\mathcal{C}$, denoted $S(\mathcal{C})$.  Here $S(\mathcal{C})=\{\rho\in \mathcal{C}^*;\, \rho(A^*A)\in [0,\infty)\, \forall\, A\in \mathcal{C},\,\rho(1)=1\}$.  If $\mathcal{C}=C(M)$, then $x(\in M) \mapsto \rho_x(\in \mathcal{C}^*)$ via $\rho_x(f)=f(x)$ for $f\in C(M)$.  Thus, $M\hookrightarrow S(\mathcal{C})\subset\mathcal{C}^*$, and the notion of distance on a metric space $M$ is extended to $S(\mathcal{C})$.  For $\sigma, \rho\in S(\mathcal{C})$: 
\[d_L(\sigma, \rho)=\sup\{|\sigma(f), \rho(f)|;\, L(f)\leq 1\}.\]

\indent Inspired by some properties of the Lipschitz seminorm on $C(M)$ for a metric space $M$, a Lip-norm is then defined as follows. 
\begin{Def} Given a semi-norm $L$ on a $C^*$ algebra, $L$ is a \textit{Lip-norm} if 
\begin{itemize}
\item $L(1)=0$ where $1$ is order unit (the identity for unital $C^*$ algebras) and\\
\item  the topology on $S(\mathcal{C})$ inherited from $d_L$ (as defined above with $L$) should coincide with the weak-$*$ topology. 
\end{itemize}
\end{Def} 

\indent With this definition of a quantum metric space, Rieffel was able to define a notion of \textit{quantum} Gromov-Hausdorff convergence of compact quantum metric spaces ( \cite{GHDQMC}), generalizing Gromov-Hausdorff convergence of metric spaces.  In \cite{GHDQMC}, it was stated that one purpose in defining quantum Gromov-Hausdorff convergence was to give a general framework for understanding assertions in physics literature of a sequence of operator algebras converging to some other operator algebra while keeping track of length information.  For example, the complex matrix algebras $M_n$ converge to the 2-sphere as $n$ goes to infinity.  See \cite{LeibQGHC}, \cite{QGHC} for further reading including a result showing $M_n$ equipped with appropriate Lip-norms converge to $C(2-sphere)$ as $n$ goes to infinity.    

\indent Despite Rieffel's definition of a compact quantum metric space producing a framework in which operator algebra convergence could be understood in a metric sense, his definition in the abelian case (considering Lip-norms on commutative $C^*$ algebras) is more general than metric spaces.  In Chapter 2, we will introduce the notion of a \textit{$W^*$-metric} on a von Neumann algebra $\mathcal{M}$ as introduced in \cite{KuWe}.  The quantum Hamming metric and classical Hamming metric are two special cases of a $W^*$-metric on the von Neumann algebras $\mathcal{B}(\mathbb{C}^{2^n})$ and $\ell^{\infty}(\mathbb{F}_2^n)$ respectively.  One nice aspect of $W^*$-metrics on $\ell^{\infty}(X)$ is that they coincide precisely with metrics on $X$.  See \cite{KuWe} for many other results including relating $W^*$-metrics, \textit{measureable metrics} and Connes' \textit{spectral triples} for Riemannian manifolds.  We mention this as Connes' \textit{spectral triples} were an original inspiration for the definition of a quantum (or non-commutative) metric space (see \cite{spectraltriple}).

\indent In \cite{KuWe} the definition of a quantum code is given, and it is a direct generalization of the definition given in Theorem 3.2 in \cite{QECC}.  In chapter 2 we will review the definition of a distance $t$ code in a $W^*$-metric space.  We'll see that a distance $t$ code for a $W^*$ metric on $\ell^{\infty}(X)$ corresponds to a subset $C\subset X$ satisfying: if $x\neq y\in C$ then the distance between $x$ and $y$ is no less than $t$.  Thus a quantum code of distance $t$ in $\ell^{\infty}(X)$ will correspond to a classical code in $X$ of distance $t$.  I'll define generalization of the \textit{recovery operator} given in \cite{QECC} for codes in finite dimensional $W^*$-metric spaces (finite dimension refers to the von Neumann algebra dimension), and then I will construct recovery operators for codes.   These results generalize known results regarding quantum error correction and classical error correction for finite metric spaces.  

\indent In Chapter 3, we introduce a certain class of $W^*$-metric spaces that come from certain representations semi-simple Lie algebras $\mathfrak{g}$, and we call them $\mathfrak{g}$-metrics (see definition \ref{g-metric}).  Then we present a method for constructing codes, and then will turn our attention to a special case of looking for codes in $W^*$-metric spaces coming from irreducible representations of $\mathfrak{su}(2)$.  One of the main results is purely an argument in convex geometry.  Tverberg's theorem in convex geometry yields a non-constructive proof of a result that we will need (see theorem \ref{Tverberg}).  As we will desire a special, constructive case of Tverberg's theorem (see theorem \ref{Tver}), we give a constructive proof of Tverberg's theorem for points in $\mathbb{R}^n$ on the \textit{moment curve}, $m(t)=(t,t^2, \ldots, t^n)$ where $t\in \mathbb{R}$.  The use of Tverberg's theorem for the construction of quantum codes detecting general subsets of operators in $\mathcal{B(H)}$ ($\mathcal{H}$ being finite dimensional) has already been seen in \cite{GNEC}.  

\indent In Chapter 4, we introduce the notion of a \textit{quantum distance distribution}.  Classically, if we are given a finite subset $C$ of a metric space $(M,d)$, then the \textit{distance distribution} for $C$ of length $t$ is defined as: 
\[B_t(C)=\frac{1}{|C|}\#\{(x,y)\in C\times C;\, d(x,y)=t\}\]

For subsets of a normed vector space that are a subgroup under addition (e.g. subspaces of $\mathbb{F}_2^n$), the above quantity is usually given the name \textit{weight distribution} or simply \textit{weight} of length $t$.  Notice for such subsets we have:
\[B_t(C)=\#\{x\in C; ||x||=t\}\]

\indent For linear codes in $C\subset \mathbb{F}_2^n$ we can define its dual code $C^{\perp}$ as the subspace of vectors orthogonal to $C$ (with respect to $\langle (x_i),(y_i)\rangle=\sum_i x_iy_i$).  The \textit{MacWilliams identity} linearly relates the weights of a code $C$ to those of its dual code $C^{\perp}$.  These identities in turn can be used to produce a linear programming problem that will yield upper bounds on the size of a code $C\subset \mathbb{F}_2^n$ of distance $t$.  

\indent A few \textit{quantum weights} have been introduced in \cite{QMac} and \cite{Rains02} for quantum codes.  In each of these works two sets of quantum weight enumerators were presented.  This is much in analogy to defining the $B_t(C)$ and $B_t(C^{\perp})$ for a subset $C$ in a metric space, and in fact for \text{additive quantum codes} this comment has literal interpretations (\cite{Rains02}).  The \textit{quantum distance distribution} we introduce is a straight forward generalization of the quantum weight $B_t$ in equation 4 of \cite{QMac}.  In fact, for the case that the $W^*$-metric corresponds to the quantum Hamming metric, the quantum distance distribution (definition \ref{quantumdd}) will be exactly the one presented in \cite{QMac}.  We'll show how the quantum distance distribution for $W^*$-metrics on $\ell^{\infty}(M)$ for $M$ finite is a generalization of the classical distance distribution.  In both \cite{QMac} and \cite{Rains02}, a linear relationship is produced between two quantum weights for the quantum Hamming metric.  Using linear programming techniques, this linear relationship can be used to establish upper bounds for the dimension of a distance $t$ quantum codes for the quantum Hamming metric.  Under this inspiration we produce a linear relationship between two analogous quantum weight distributions (one of them being the quantum distance distribution) for $W^*$-metrics coming from $\mathfrak{su}(2)$ representations.  We utilize the \textit{Wigner $6j$-symbols} to make this relationship.  In turn, we can also construct a linear programming problem to establish an upper bound for the dimension of a distance $t$ code for a fixed $W^*$-metric space. 
\end{section}

\begin{section}{Review of Quantum Codes and Quantum Error Correction}
\begin{subsection}{Quantum Probability}
\indent\indent In this section we review some basic concepts of quantum probability and quantum operations.  These are the quantum analogues of classical probability and Markov maps.  We will then continue our discussion into error correcting codes and quantum error correction.  For a more in depth discussion see \cite{NeCh}, \cite{CQinfo},\cite{QECC} or \cite{OpAlg}.

\indent A \textit{quantum random variable algebra} is a von Neumann algebra $\mathcal{M}$.  All classical random variable algebras are commutative von Neumann algebras, and they can all be identified as $L^{\infty}_\mu(M)$ for a measure space $M$ with measure $\mu$.  The set of \textit{positive elements} of $\mathcal{M}$, denoted $\mathcal{M}_{+}$, are all elements that can be written as $XX^*$ for $X\in\mathcal{M}$.  A \textit{state} on $\mathcal{M}$ is an element $\rho$ in the dual space of $\mathcal{M}$, $\mathcal{M}^*$, that is non-negative on the positive elements of $\mathcal{M}$.  A \textit{normalized state} $\rho$ is a state such that $\rho(1)=1$.  The set of self adjoint projections in $\mathcal{M}$ are considered boolean random variables (a projection $P$ satisfies $P^2=P$).  For commutative $\mathcal{M}=L^{\infty}_{\mu}(M)$ projections are in one-to-one correspondence with measurable subsets of $M$.  Thus if $P$ is a projection corresponding to $S\subset M$ and $\rho$ is a normalized state on $L^{\infty}_{\mu}(M)$, then $\rho(P)$ can be interpreted as the probability that $\rho$ is in $S$.  In general, we say the expectation value of a random variable $X$ relative to a state $\rho$ is $E_{\rho}(X)=\rho(X)$.  

\indent Since we will be mostly only considering finite dimensional $\mathcal{M}$, we can make an identification between $\mathcal{M}^*$ and $\mathcal{M}$.  The classification theorem for finite dimensional von Neumann algebras states that $\mathcal{M}\simeq\bigoplus_i M_{n_i}(\mathbb{C})$.  Here $M_{n}(\mathbb{C})$ is a complex matrix algebra of dimension $n^2$.  The Hilbert-Schmidt (HS) form on $\mathcal{M}$ is defined as $(X,Y)=(\oplus_i X_i, \oplus_i Y_i)=\sum_i Tr(X_i^*Y_i)\defeq Tr(X^*Y)$.  The HS form gives $\mathcal{M}$ the structure of a Hilbert space, and thus $\rho\in \mathcal{M}^*$ can be identified with an element $X_\rho\in\mathcal{M}$ via $\rho(\cdot)=(X_\rho, \cdot)$ by the finite dimensional form of the Riesz representation theorem.  States on $\mathcal{M}$ are thus identified with positive elements of $\mathcal{M}$, and normalized states are ones where $(X_\rho, I)=1$ or we can just say $Tr(X_\rho)=1$. 

\indent The time evolution of a quantum system in most of quantum information theory is assumed to obey the Schr\"odinger picture.  It is assumed that the quantum system is closed and is non-relativistic.  For a state $\rho\in\mathcal{B(H)}$, its time evolution is determined by a collection of unitary operators in $\{U_t\}\subset\mathcal{B(H)}$.  At time $t$, the state $\rho$ becomes $U_t\rho U_t^{-1}$.  

\indent We may also make a measurement on a quantum system with a boolean random variable $P\in\mathcal{B(H)}$.  We know the probability of observing $P$ is $\rho(P)$ (we can say $Tr(\rho P)$ for finite dimensional systems).  If $P$ was in fact observed (so certainly $\rho(P)\neq 0$), the post measurement state is: 
\begin{equation} \frac{\rho(P\,\cdot\,P )}{\rho(P)}.\end{equation}

If we made observations of $\rho$ with many booleans $\{P_i\}$ where $P_iP_j=\delta_{ij}P_i$ and $\sum_i P_i=I$ (a complete measurement where some $i$ will be observed), then we observe $i$ with probability $\rho(P_i)$.  If we are in a memoryless system where no record of the measurement outcome (but we do know there was one), then the state $\rho$ is in classical superposition of possible outcomes:
\[\sum_i \rho(P_i\,\cdot),\]
or if we use the picture that $\rho\in\mathcal{B(H)}$
\[\sum_i P_i\rho P_i.\]
 
\indent  A quantum operation is a map on a quantum state that is meant to model the evolution of a quantum state with the possible presence of some interaction from the environment.  Thus if our quantum system we wish to make observations on is based on a Hilbert space $H$ and the environment is modeled by a Hilbert space $E$, then the entire quantum system is $H\otimes E$.  If we assume unitary time evolution of the entire system $H\otimes E$ and yet only care about the evolution within $H$ of a state $\rho$ originally in $\mathcal{B(H)}$, then at any point in time the state will evolve to a state of the form (for finite dimensional $H$): 
\[ \sum_i A_i\rho A_i^*.\]
Here $A_i\in\mathcal{B(H)}$, and $\sum_i A_i^*A_i\leq I$.  If no part of the total quantum system and environment is destroyed, then $\sum_i A_i^*A_i= I$.  The operators $A_i$ come from mixing both unitary evolution of the environment with \textit{memoryless} boolean observations of the environment.  Memoryless boolean measurements here means we are taking boolean measurements on the environment, but we didn't make an observation to see what state the post measurement system was in.  Thus, the post measurement system is in a classical probabilistic superposition of possible out come states.  We may notice that a map $\mathcal{A}(\cdot)=\sum_i A_i\cdot A^*_i$ sends positive elements to positive elements.  In fact $\mathcal{A}$ is \textit{completely positive} (CP) since $\mathcal{A}\otimes I\in\mathcal{B(H\otimes V)}$ for \textit{any} Hilbert space $\mathcal{V}$ sends positive elements to positive elements.  Choi's theorem (\cite{choi}) identifies all completely positive maps on finite dimensional $\mathcal{B(H)}$ as those in the form of $\mathcal{A}$ for some collection of operators $\{A_i\}$.  If a map $\mathcal{A}$ is completely positive and preserves probability, then we call it completely positive trace preserving (CPTP).
\end{subsection}

\begin{subsection}{Quantum Codes}
\indent\indent We next move to reviewing quantum codes and quantum error correction.  For the remainder of this discussion we will assume $\mathcal{H}$ is a finite dimensional Hilbert space.  Suppose we have a collection of operators $E\subset\mathcal{B(H)}$ where quantum operations of the form:
\[\sum_i E_i\cdot E^*_i\]
are viewed as noise in the system.  If the set $\{E_i\}\subset E$, then we will call such operators \textit{errors} from $E$.  
\begin{Def} If $P$ is a projection in $\mathcal{B(H)}$ where $PXP\propto P$ for all $X\in E$, then we say $P$ \textit{detects} noise from $E$, and $P$ is a \textit{code projector}.
\end{Def}
By the linearity of the noise detection condition for $P$, we know that $P$ also detects noise from the $*$-closed linear span of $E$.    

\begin{Def}  If $E$ is $*$-closed linear subspace of $\mathcal{B(H)}$ and $P$ detects errors from $E^2$ (the linear span of $xy$ for $x,y\in E$), then we say the code $P$ \textit{corrects} errors from $E$. 
\end{Def}

\indent If a code corrects errors from $E$, then $PX^*YP=(X,Y)P$ for $X,Y\in E$ and $(X,Y)\in\mathbb{C}$.  It is easy to check that $(\cdot, \cdot)$ is a (possibly degenerate) Hermitian form on $E$.  Thus quotienting $E$ by the kernel of $(\cdot, \cdot)$ (we again denote this as $E$) yields a Hilbert space.  Thus if $C$ is the support of the code projector $P$, we can form a Hilbert space $E\otimes C$ in the usual sense of the tensor product of two finite dimensional Hilbert spaces.  We then have the following theorem.
\begin{thm} \label{firstembed}The Hilbert space $E\otimes C$ isometrically embeds into $\mathcal{H}$ via $x\otimes v\mapsto x(v)$. 
\end{thm}
\begin{proof}  We define an embedding of $E\otimes C\hookrightarrow \mathcal{H}$ via $X\otimes v\mapsto Xv$.  For $X,Y\in E$ and $v,w\in C$, we now only check 
\[\langle X\otimes v, Y\otimes w\rangle= (X,Y)\langle v, w\rangle=\langle v, (X,Y)w\rangle=\langle v,X^*Yw\rangle=\langle Xv, Yw\rangle.\]
Now since this equality holds on simple tensors, by linearity it holds for all elements of $E\otimes C$.  
\end{proof}
In essence this theorem says there is enough orthogonal room in $\mathcal{H}$ for $E$ to move the subspace $C$ so that we can distinguish which error occurred.  Now we can give better justification for the title ``correctable errors" in the following.  

\indent We say a CPTP map $\mathcal{A}$ is an \textit{error correcting transformation} for $E$ if for any error from $E$, say $\mathcal{E}$, we have $\mathcal{A}\circ \mathcal{E}(\rho)\propto \rho$ for any state 
$\rho$ where $P\rho P=\rho$.  We can now state the following theorem. 

\begin{thm} \label{errorcorrection} A quantum code detects errors from $E^2\subset \mathcal{B(H)}$ iff an error correcting transformation exists for errors from $E$.  
\end{thm}  
\begin{proof} Using theorem \ref{firstembed} we can see errors from $E$ sends states supported on $C$ to (perhaps un-normalized) states on $E\otimes C$.  We recall the definition of the \textit{partial trace} operator.  We consider its' action on simple tensors in $A\otimes B\in\mathcal{B}(E\otimes C)=\mathcal{B}(E)\otimes\mathcal{B}(C)$ (here we assume both $E$ and $H$ are finite dimensional).  The partial trace is defined as $Tr_{C}(A\otimes B)=Tr(A)B\in \mathcal{B}(C)$.  Then $Tr_{C}$ is defined by extending linearly to non-simple tensors.  The partial trace operator here is well defined since (one can check) it is dual to the embedding $\mathcal{B}(C)\rightarrow \mathcal{B}(C\otimes H)$ via $A\mapsto A\otimes 1$ where $1$ is the identity in $\mathcal{B}(E)$.  We also state that the partial trace operator is a completely positive, trace preserving map.  

\indent Let $\mathcal{O}$ be an CPTP map from the operators supported on the orthogonal complement of $E\otimes C\subset \mathcal{H}$ to operators on $C$ (here we are using theorem \ref{errorcorrection}).  It is easy to check that $Tr_{C}\oplus \mathcal{O}$ is an error correcting transformation for errors from $E$.   
\end{proof}

\indent We end this section by mentioning a result in \cite{GNEC} that shows the existence of error detecting codes.  Given a Hilbert space $\mathcal{H}$ of dimension $N$, if the dimension of a set of errors $E\subset\mathcal{B(H)}$ is $M$, then there exists a code of dimension at least $\lceil \frac{N}{M}\rceil \frac{1}{M+1}$.  
\end{subsection}

\begin{subsection}{Quantum Hamming Filtration and Code Distance}
\indent\indent We begin by introducing a \textit{qubit}.  A qubit is the quantum analogue of the classical two state probabilistic bit; where a state is probabilistic sum of states $[0],[1]\in\mathbb{F}_2$.  A qubit is a state in $\mathcal{B}(\mathbb{C}^2)$.  Thus any qubit can be written as a probabilistic sum of two one-dimensional, self-adjoint projections $P_1$ and $P_2$ where $P_1P_2=0$.  A string of n-qubits is a state in $\bigotimes_{i=1}^n \mathcal{B}(\mathbb{C}^2)$; that being analogous to a string of n-bits being states on $\Pi_{i=1}^n \{[0],[1]\}=\mathbb{F}_2^n$.  

\indent Many models for a quantum computer are based on strings of qubits.  Then, the severity of an error in a quantum computer is based on how many qubits the error affected.  Analogously, the severity of errors in a classical computer is assessed by the number of bits affected.  In actuality, most of the time we can not guarantee errors will only affect a fixed number of qubits, but we can approximate errors by ones only affecting a fixed number of qubits if the noise affecting the system is not too intense.  There are many articles on this subject, but we can suggest a discussion in \cite{CQinfo}.

\indent We will introduce the multi-Pauli operators and present how they are used to give a notion of distance to errors on strings of n-qubits.  The multi-Pauli operators are a basis for $\bigotimes_{i=1}^n \mathcal{B}(\mathbb{C}^2)$.  A multi-Pauli operator is any operator of the form:
\[\sigma_{i_1}\otimes\sigma_{i_2}\otimes\ldots\otimes \sigma_{i_n}.\]
Here each $\sigma_{i_k}$ is the identity or one of a set of self adjoint operators called the Pauli operators:
\begin{equation} \sigma_1=
\begin{bmatrix}
0 & 1\\
1 & 0
\end{bmatrix}
\qquad \sigma_2 = 
\begin{bmatrix}
0 & -i\\
i & 0
\end{bmatrix}
\qquad \sigma_3 =
\begin{bmatrix}
1 & 0\\
0 & -1
\end{bmatrix}
\end{equation}

\indent Notice $\sigma_1\sigma_2=i\sigma_3$, $\sigma_1^2=\sigma_2^2=\sigma_3^2=1$ and knowing the Pauli operators are self adjoint imply all multi-Pauli matrices with coefficients $\pm 1, \pm i$ form a finite group called the multi-Pauli group.  A multi-Pauli operator with exactly $t$ tensor terms not equal to the identity is said to be a distance $t$ operator.  The span of the distance $\leq t$ multi-Pauli operators we can denote as $\mathcal{E}_{t}$, and in turn this yields a filtration on $\bigotimes_{i=1}^n \mathcal{B}(\mathbb{C}^2)$ called the \textit{quantum Hamming filtration}.  The filtration satisfies:
\begin{enumerate}
\item $\mathbb{C}I=\mathcal{E}_0\subset \mathcal{E}_1\subset\ldots\subset \mathcal{E}_n=\bigotimes_{i=1}^n \mathcal{B}(\mathbb{C}^2)$,\\
\item $\mathcal{E}_s\cdot \mathcal{E}_t\subset \mathcal{E}_{s+t}$ and \\
\item $\mathcal{E}_t^*=\mathcal{E}_t$.  
\end{enumerate}

\indent We can use the above filtration to define a distance for quantum operations on n-qubits.  A quantum operation $\Phi$, by Choi's theorem, can be written as
\begin{enumerate}
\item $\Phi(\cdot)=\sum_j E_j \cdot E_j^*$ where\\
\item $\sum_j E_j^* E_j \leq I$.
\end{enumerate}

The operators $E_j$ are not unique to the representation of $\Phi$, but non-the-less for finite dimensions $\min\{t:\,E_i\in \mathcal{E}_t\,\forall i\}$ is unique to $\Phi$.  Thus an error operator where $s=\min\{t:\,E_i\in \mathcal{E}_t\,\forall i\}$ is said to be an error on $s$ qubits.  This result is an easy corollary from the following lemma.

\begin{lem} If the operators $\{E_i\}\subset \mathcal{B}(V)$ are linearly independent, then the superoperators $E_i\cdot E_j^*$ $\forall i,j$ are linearly independent.  
\end{lem}
\begin{proof} 
 We first recall that if $B_1$ and $B_2$ are bases for finite dimensional vector spaces $V_1$ and $V_2$, then $\{X\otimes Y;\,X\in B_1,\,Y\in B_2\}$ is a basis for $V_1\otimes V_2$.  Also, if $\{E_i\}$ is a basis of $\mathcal{B}(V)$, then so is $\{E_i^*\}$.  Thus in $\mathcal{B}(V)\otimes\mathcal{B}(V)$, the operators $E_i\otimes E_j^*$ are linearly independent.  The map $\mathcal{B}(V)\otimes\mathcal{B}(V)\rightarrow\mathcal{B}(\mathcal{B}(V))$ defined by $A\otimes B\mapsto A\cdot B$ extended linearly is surjective.  One can check this by beginning with a basis $\{|v_i\rangle\}\subset V$, and then check that the operators $|v_i\rangle\langle v_j|\cdot |v_k\rangle\langle v_l|$ are a basis of $\mathcal{B}(\mathcal{B}(V))$.  Thus this map is an isomorphism.  The result now follows.
\end{proof}
\begin{lem} If $\sum_i E_i\cdot E_i^*=\sum_j F_j\cdot F_j^*$ as superoperators, then the subspaces of $\mathcal{B}(V)$ spanned by $\{E_i\}$ and $\{F_j\}$ are the same.  
\end{lem}
\begin{proof} Let $\mathcal{E}$ and $\mathcal{F}$ be the subspaces of $\mathcal{B}(V)$ spanned by $\{E_i\}$ and $\{F_j\}$ respectively.  Let $B$ be a basis of $\mathcal{E}$ and $B'$ be a basis such that $B\cup B'$ is a basis of $\mathcal{B}(V)$.  If we express $\sum_iE_i\cdot E_i^*$ in terms of elements of $B$ and $\sum_j F_j\cdot F_j^*$ in terms of elements from $B\cup B'$, then by the previous lemma these expressions should coincide as expressed in linear sums of operators $X\cdot Y^*$ where $X,Y\in B$.  Thus $\mathcal{F}\subset \mathcal{E}$.  By reversing the roles of $\{E_i\}$ and $\{F_j\}$ in the previous argument, we see $\mathcal{E}\subset \mathcal{F}$.  
\end{proof}
\begin{cor} If $\Phi(\cdot)=\sum_j E_j \cdot E_j^*$, then $s=\min\{t:\,E_i\in \mathcal{E}_t\,\forall i\}$  is unique to $\Phi$.  
\end{cor}

\indent Thus, if a code projection $P\in \bigotimes_{i=1}^n \mathcal{B}(\mathbb{C}^2)$ detects errors from $\mathcal{E}_{t+1}$ we say it is a \textit{distance $t$ code}.  By the filtration condition, a distance $t$ code corrects errors from $\mathcal{E}_{\lfloor \frac{t+1}{2}\rfloor}$.  We know from theorem \ref{errorcorrection} that a single error correcting transformation, for a distance $t$ code, exists for any distance $\lfloor \frac{t+1}{2} \rfloor$ errors.    

\indent A class of quantum distance $t$ codes have been constructed called \textit{additive} or \textit{stabilizer} codes.  These codes all correspond to eigen value $1$ subspaces of abelian subgroups of a multi-Pauli group.  In \cite{CodesExist} and \cite{QECorthgeo} for example, such codes are constructed, and lower bounds for the dimension of minimal distance $t$ codes are given.  These lower bounds are far better than the general bound mentioned earlier in \cite{GNEC}.  In \cite{GF(4)codes} a correlation is made between these codes and certain types of self-dual additive codes in vector spaces over $GF(4)$.  This result makes is possible to apply classical coding techniques to the arena of quantum codes.  Also, quantum codes not fitting into the class of additive or stabilizer codes have been found.  For example in \cite{nonadditive} a distance two, 6-dimensional code projection is found in the collection of five qubits.  This code is known to be better than any additive/stabilizer code.  
\end{subsection}
\end{section}

   \chapter[%
      Short Title of 2nd Ch.
   ]{%
     $W^*$-metric Spaces
   }%
   \label{ch:2ndChapterLabel}
   \begin{section}{$W^*$-metric: Definitions, Examples}
\indent
\indent In \cite{GNEC}, one generalization of the quantum Hamming filtration was given.  In that article, $\mathcal{A}\subset \mathcal{B(H)}$, for finite dimensional Hilbert space $\mathcal{H}$, is an \textit{interaction algebra} iff $\mathcal{A}$ is a $*$-closed, unital sub-algebra (i.e. a finite dimensional von Neumann algebra) with a filtration specified by the following.  We begin with a $*$-closed linear subspace, $\mathcal{J}_1$ of  $\mathcal{A}$ containing the identity.  Then the filtration is given by $\mathcal{J}_1\subset\mathcal{J}_1^2\subset\ldots\subset \mathcal{J}_1^k=\mathcal{A}$.  Here $\mathcal{J}_1^d$ is the linear span of the product of no more than $d$ operators from $\mathcal{J}_1$.  Clearly the quantum Hamming filtration is a special case of an interaction algebra with $\mathcal{A}=\bigotimes^{n}\mathcal{B}(\mathbb{C}^2)$ and $\mathcal{J}_1$ is the span of multi-Pauli operators with only one tensor term not equal to the identity.  

\indent A distance $t$ code is then a projection $P\in\mathcal{B(H)}$ that satisfies the usual quantum error correction condition for operators in $\mathcal{J}_{t-1}$.  By this we mean $PEP\propto P$ for all $E\in \mathcal{J}_{t-1}$.  As in \cite{GNEC}, we recall a theorem regarding $*$-closed algebras of operators on finite dimensional Hilbert spaces.

\begin{thm}\label{finiteVN} Let $\mathcal{E}$ be a finite dimensional von-Neumann algebra acting on a Hilbert space $\mathcal{H}$.  Then $\mathcal{H}$ is isomorphic as a Hilbert space to the direct sum, 
\[\mathcal{H}\approx\oplus_i \mathcal{C}_i\otimes \mathcal{Z}_i\]
where $\mathcal{C}_i$ and $\mathcal{Z}_i$ are Hilbert spaces.  Here this significance of $\{\mathcal{C}_i\}$ and $\{\mathcal{Z}_i\}$ are that the von-Neumann algebra $\mathcal{E}=\oplus_i  \mathcal{I}^{\mathcal{C}_i}\otimes Mat(\mathcal{Z}_i)$, and the commutant $\mathcal{E}'=\oplus_i Mat(\mathcal{C}_i)\otimes \mathcal{I}^{\mathcal{Z}_i}$. 
\end{thm}

\indent As noticed in \cite{GNEC}, if $\mathcal{E}$ (as above) are a collection of errors, then states in $\mathcal{C}_i$ are inherently protected from errors.  They call the subsystems $\mathcal{C}_i$  \textit{noiseless subsystems}.  We will make mention that in Theorem 5 of \cite{GNEC}, they prescribe an equivalence between noiseless subsystems (through correcting errors \textit{before} they occur) and distance $t$ quantum codes.  We will come back to a discussion of noiseless subsystems in the context of $W^*$-metric spaces, but first we will discuss a relationship between interaction algebras and graph metrics.  

\indent A graph $\Gamma=(V,E)$ is a set of vertices $V$ and edges $E$ which are specified by the pair of vertices each edge connects.  The graph metric on $\Gamma$ begins with specifying each pair of points in $E$ is distance 1 away from each other.  Then distances between other points is given by the shortest edge path connecting them (if no such path exists, then the distance is said to be infinite). 

\indent We will briefly describe a ``toy" construction of an interaction algebra from a graph $\Gamma$ with graph metric $(\cdot, \cdot)$.  We first consider a Hilbert space, $H_{\Gamma}$, generated by the set of vertices $\{v_i\}$ in a graph $\Gamma$ (we'll denote the normalized vector representatives of vertices by $\{|v_i\rangle\}$).  We then form a $*$-invariant subspace of $B(H_{\Gamma})$, $\mathcal{J}_1=\mathrm{span}\{|v_i\rangle\langle v_j|:\, (v_i,v_j)\leq1\}$.  Notice the identity is in $\mathcal{J}_1$ as it should be for interaction algebras.  The identity operator is somehow a ``length zero" operator, but it is never directly considered that in \cite{GNEC}.  In the graph metric scenario above it seems natural to let the ``length zero" operators, say $\mathcal{E}_0$, be the span of $|v_i\rangle\langle v_i|$; this subspace is isomorphic to $\ell^\infty(vert(\Gamma))$.  We could similarly define $\mathcal{E}_d\defeq \mathrm{span}\{|v_i\rangle\langle v_j|:\,(v_i,v_j)\leq d\}$.  Then it happens to be that $\mathcal{E}_d=\mathcal{E}_1^d$ for, $d\geq 1$, as for interaction algebras. We now present  a possible operator algebra generalization of a metric space; which in turn is another approach to non-commutative or quantum metric spaces. 

\begin{Def} (\cite{KuWe})  A \textit{W$^*$-filtration} on the bounded operators on a Hilbert space $\mathcal{H}$, $\mathcal{B(H)}$,  is a set of weak operator closed subspaces $\{\mathcal{E}_t:\,t\in\mathbb{R}_{\geq 0}\}$ satisfying:
\begin{enumerate}
\item The identity is in $\mathcal{E}_{t}$ for all $t$,\
\item $\forall t\quad\mathcal{E}_t=\mathcal{E}_t^*$,\
\item $\mathcal{E}_{t}\mathcal{E}_s \subseteq \mathcal{E}_{s+t}$,\
\item $\mathcal{E}_{t}=\bigcap_{s>t}\mathcal{E}_{s}$.\
\end{enumerate}
\end{Def}

 \indent Take notice that the $\mathcal{E}_t$ are bimodules over $\mathcal{E}_0$.  We notice that $\mathcal{E}_0$ is a weak operator closed, $*$-closed algebra containing the identity.  From definition \ref{vonNeumann}, we see $\mathcal{E}_0$ is a von Neumann algebra.  Also from definition \ref{vonNeumann}, we see that the commutant of $\mathcal{E}_0$ is a von Neumann algebra.

\indent We will also mention a few more definitions regarding $W^*$-filtrations. The \textit{open filtration term} of degree $t$, denoted $\mathcal{E}_{<t}$, is $\cup_{s<t} \mathcal{E}_s$.  The \textit{pure filtration term} of degree $t$ , denoted $\mathcal{E}_{=t}$ is the quotient $\mathcal{E}_t/\mathcal{E}_{<t}$.   For $W^*$-metrics on finite dimensional von Neumann algebras in $\mathcal{B(H)}$ (for finite dimensional $\mathcal{H}$), we can decompose $\mathcal{B(H)}=\mathcal{E}_0\oplus \bigoplus_i \mathcal{E}_{=\alpha_i}$ using the Hilbert-Schmidt Hermitian form to embed $\mathcal{E}_{=\alpha}\hookrightarrow\mathcal{B(H)}$.  
 
\indent We can now define a \textit{$W^*$-metric} on a von Neumann algebra $\mathcal{M}\subset \mathcal{B(H)}$ as a $W^*$-filtration on $\mathcal{B(H)}$ such that the zero term, $\mathcal{E}_0$, is the commutant of $\mathcal{M}$.

\indent We won't be deeply addressing any $W^*$-metrics on infinite dimensional von Neumann algrebras, but the definition is a very fruitful one in the infinite dimensional case as well (see \cite{KuWe}).  We will only be directly considering $W^*$-metrics on finite dimensional von Neumann algebras as they may be useful in quantum information theory.  In the following, we outline a scenario to describe some potential interpretive use of $W^*$-metrics.  Although, it is not the aim of this presentation to relate this to a real world model for quantum computation or quantum information theory. 

\indent Let's begin by considering a collection of operators $\mathcal{E}\subset \mathcal{B(H)}$ where quantum operations of the form $\mathcal{O}(\rho)=\sum_{i} E_{i} \rho E_{i}^{*}$, for $E_i \in\mathcal{E}$, are somehow completely unprotectable.  Perhaps it is constant noise that would be too costly to protect states from.  We in turn will look for states who are inherently protected from noise coming from any quantum operations arising from the von Neumann algebra generated from $\mathcal{E}$.  We denote this algebra as $\mathcal{E}_0$.  We can consider $\mathcal{E}_0$ as an interaction algebra with one filtration term, namely the entire algebra.  If we decompose $\mathcal{H}$ as in \ref{finiteVN} (considering $\mathcal{E}_0$ as $\mathcal{E}$ there) then we see states in $\mathcal{E}_0'$ are states in noiseless subsystems that are inherently protected from noise in $\mathcal{E}_0$.    

\indent We digress for a moment taking an alternate view considering $\mathcal{E}_0'$ as the random variable algebra that we make measurements on \textit{any} state with.  If $\rho$ is any state acted upon by an error of the form $\sum_i E_i\rho E^*_i$ for $E_i\in \mathcal{E}_0$, then the expectation value of $A\in \mathcal{E}'_0$ with respect to an error distorted state has an outcome of the form:
\[Tr(\sum_i E_i\rho E^*_i A)=Tr(\sum_i E^*_iE_i\rho A).\]
If the error is a trace preserving map, then the outcome of the measurement is completely unchanged by the error (since $\sum_i E^*_iE_i=I$).  Otherwise (for $\sum_i E^*_iE_i<I$), the error only affects the outcome of a measurement from $\mathcal{E}_0'$ as much as information about the existence of the state was lost in the error.  

\indent We now continue our previous point regarding $\mathcal{E}_0$ as a von Neumann algebra generated by errors we have no control over (i.e. we take no active role in error correction).  We know for error distorted states that we can still reliably find the expectation values of random variables $\mathcal{E}_0'$ with respect to $\rho$.  We may also have another set of errors from a $*$-closed subspace $\mathcal{E}_1$ that we would \textit{actively} want to protect our system from.  One could imagine a scenario where we would want to balance the amount of quantum entanglement of states used to store information with the type of errors possible (i.e. we want to reliably use all of $\mathcal{E}_0'$ as a random variable algebra).  Since the constant noise from $\mathcal{E}_0$ would be happening amongst errors from $\mathcal{E}_1$, completely positive maps resulting from errors from $\mathcal{E}_1$ should include pre and post errors from $\mathcal{E}_0$.  Thus the summands of such quantum operations would be of the form $(e_1Ee_2) \rho (e_1Ee_2)^*$ with $E\in \mathcal{E}_1$, $e_1,e_2\in \mathcal{E}_0$ and $\rho$ is a state.  Thus, we would be performing error correction on the entire $\mathcal{E}_0$ bimodule generated by $\mathcal{E}_1$.  Beyond a discussion strictly focused on metric spaces, that is one reason for wanting the $W^*$-filtration terms to be bi-modules over the 0-term $\mathcal{E}_0$.  

\indent We will come back to discussing error correction.  For now, we wish to make mention of an elementary but important example of how $W^*$-metrics are related to classical metric spaces. 

\begin{subsection}{$W^*$-metrics on $\ell^{\infty}(M)$ and Metrics on $M$}
\indent\indent We will make an important comment, necessary only here in our discussion, that in \cite{KuWe} it is proved that the class of $W^*$-metrics on a von Neumann algebra $\mathcal{M}$ is independent of the faithful representation of $\mathcal{M}$ on a Hilbert space.  Thus we only will be using the representation of $\ell^{\infty}(M)$ as multiplication operators on $\ell^2(M)$.  

\indent Previously we constructed a $W^*$-metric space from a graph, but a similar construction can be done to any metric space $(M,d)$.  This is done in a much more general setting of \textit{finitely decomposable} measure spaces in \cite{KuWe}, but we only present a purely atomic example.  We begin with the Hilbert space $\ell^2(M)$, the square integrable functions on $M$ with the atomic measure.  Then we say $A\in \mathcal{E}_t\subset \mathcal{B}(\ell^2(M))$ iff $\chi(p)A\chi(q)=0$ for all $p,q\in M$ such that $d(p,q)>t$ ($\chi(p)$ is the characteristic function of $p\in M$).  One can check that this defines a $W^*$-metric space with $\mathcal{E}_0'=\ell^{\infty}(M)$.  Reversing the construction, assume we are given a $W^*$-metric space where $\mathcal{E}_0'=\ell^{\infty}(M)$.  We can construct a metric on $M$ by defining $d(p,q)=\inf\{t: \exists  A\in \mathcal{E}_t\,\,\chi(p)A\chi(q)\neq 0\}$.  Since the argument isn't difficult and gives an example of how to understand $W^*$-metric spaces, we will present it here. 
\begin{thm} $W^*$-metrics on $\ell^{\infty}(M)$ are equivalent to metrics on $M$.
\end{thm}
\begin{proof}   We begin with metric on $M$.  We first show that $A\in \mathcal{E}_t$ iff $\chi(p)A\chi(q)=0$ for all $p,q\in M$ such that $d(p,q)>t$ defines a $W^*$-filtration with $\mathcal{E}_0'=\ell^{\infty}(M)$.  Clearly, $\mathcal{E}_t$ is a $*$-invariant linear subspace.  If $A_n\in \mathcal{E}_t$ is a sequence converging in the weak operator topology to $A$, then for all $d(p,q)>t$ 
\[\langle \chi(p),A_n \chi(q)\rangle=\langle \chi(p), (\chi(p)A_n\chi(q))\chi(q)\rangle\rightarrow\langle \chi(p), (\chi(p)A\chi(q))\chi(q)\rangle =0.\]
Thus, $\chi(p)A\chi(q)=0$, and we have that $\mathcal{E}_t$ is weak operator closed. 

\indent Now if $A\in \mathcal{E}_t$ and $B\in \mathcal{E}_s$, suppose there exists $d(p,q)>s+t$ such that $\chi(p)AB\chi(q)\neq 0$.  Then, $\chi(p)A\sum_{a\in M}\chi(a)B\chi(q)\neq 0$.  Thus, there is some $a\in M$ such that \\ $(\chi(p)A\chi(a))(\chi(a)B\chi(q))\neq 0$.   It follows $\chi(p)A\chi(a)\neq 0$ and $\chi(a)B\chi(q)\neq 0$, but that means $d(p,a)\leq t$ and $d(a,q)\leq s$.  This contradicts $d(p,q)> s+t$.  Also, $\cap_{s>t}\mathcal{E}_s=\mathcal{E}_t$ follows directly from the definition of $\mathcal{E}_t$.   Lastly, clearly $\ell^{\infty}(M)\subset\mathcal{E}_0$.  If $A\in \mathcal{E}_0$, then $\langle \chi(p),A\chi(q)\rangle\neq 0$ iff $p=q$.   This implies $\chi(p)A=A\chi(p)$ for all $p\in M$.  Since $\ell^{\infty}(M)$ is a maximal abelian sub-algebra, $A\in\ell^{\infty}(M)$.   Thus $\mathcal{E}_0=\ell^{\infty}(M)=\mathcal{E}_0'$.

\indent For the other direction, we begin with a $W^*$-filtration in which $\mathcal{E}_0'=\ell^{\infty}(M)$, and we define $d(\cdot, \cdot)$ as above.  Since the identity is in $\mathcal{E}_0$, $d(p,p)=0$.  Also, if $A\in \mathcal{E}_0=(\ell^{\infty}(M))'$, then $\chi(p)A\chi(q)=A\chi(p)\chi(q)\neq 0$ implies $p=q$.  Thus, $d(p,q)=0$ iff $p=q$.  Symmetry of $d$ follows from $*$-invariance of $\mathcal{E}_t$.  The triangle inequality $d(p,q)\leq d(p,r)+d(r,q)$ can be demonstrated by the following.  Suppose $d(p,r)=s$ and $d(r,q)=t$.  Thus $\forall \epsilon >0\, \exists$ $A_{s+\epsilon}\in \mathcal{E}_{s+\epsilon}$ and $B_{t+\epsilon}\in \mathcal{E}_{t+\epsilon}$ where $\chi(p)A_{s+\epsilon}\chi(r)\neq 0$ and $\chi(r)B_{t+\epsilon}\chi(q)\neq 0$.  Since $\chi(\cdot)$ are rank 1 projections, $\forall \epsilon$ $\chi(p)A_{s+\epsilon}\chi(r)B_{t+\epsilon}\chi(q)\neq 0$.  Thus $\forall \epsilon$, $d(p,q)\leq d(p,r)+d(r,q)+2\epsilon$, since $A_{s+\epsilon}\chi(r)B_{t+\epsilon}\in \mathcal{E}_{s+t+2\epsilon}$.  This yields the triangle inequality.  
\end{proof}
\end{subsection}
\end{section}

\begin{section}{Codes in $W^*$-metric Spaces}

\indent\indent In Chapter 1, we reviewed the definition of quantum code $P\in \mathcal{B(H)}$ that detects errors $\mathcal{E}\subset\mathcal{B(H)}$.  From the error detection condition we know $P$ detects errors from the $*$-closed linear subspace generated from $\mathcal{E}$, which from here on we will denote again as $\mathcal{E}$.  We also know that if $P$ in fact detected errors from $\mathcal{E}^2$ then this is equivalent to there existing an error correcting transformation for states supported on $P$.  

\indent  In this section we will generalize these results for any finite dimensional $W^*$-metric space.  We define what an error detecting code is for any $W^*$-metric space (including infinite dimensional von Neumann algebras).  For finite dimensional $W^*$-metric spaces we will define an $\mathcal{E}_0$-error correcting transformation where $\mathcal{E}_0$ is the zero term in a $W^*$-filtration.  When  $\mathcal{E}_0'=\mathcal{B(H)}$ all usual notions of quantum codes and quantum error correction result.  Also when $\mathcal{E}_0'=\ell^{\infty}(M)$ we will arrive at classical notions of codes and  error correction.  

\begin{subsection}{Definitions and Theorems}


\indent\indent Given a $W^*$-metric on a von Neumann algebra $\mathcal{M}\subset \mathcal{B(H)}$, let $\{\mathcal{E}_t\}_{t\geq 0}$ be the $W^*$-filtration terms yielding the $W^*$-metric on $\mathcal{M}$.  We define a \textit{distance $t$} quantum code in $\mathcal{M}$ to be a projection $P\in \mathcal{M}$ such that:
\[ P\mathcal{E}_{<t}P=\mathcal{E}_0P.\]

\indent We can now use the fact that $\mathcal{E}_{<t/2}\cdot \mathcal{E}_{<t/2}\subset \mathcal{E}_{<t}$ to define a $\mathcal{E}_{0}P$-valued sequilinear operator on $\mathcal{E}_{<t/2}$.  The form is defined as $(E, F)=PE^{*}FP\in \mathcal{E}_0P$.  The subspace $\mathcal{E}_{<t/2}P$ is a right $P\mathcal{E}_{0}$-module, and it has a quotient that is an \textit{inner product $P\mathcal{E}_0$-module}.  We review the definition.

\begin{Def} Let $\mathcal{C}$ be a $C^*$-algebra.  An \textit{inner product $\mathcal{C}$-module} is a complex vector space $V$ with a right $\mathcal{C}$ action and a map $(\cdot,\cdot):V\times V\rightarrow \mathcal{C}$ which satisfies the following:
\begin{enumerate}
\item $\mathbb{C}$-linear in the right argument,\
\item right $\mathcal{C}$-linear in the right argument, i.e. $(v,wC)=(v,w) C$ for $C\in \mathcal{C}$,\
\item $(v,w)^{*}=(w,v)$,\
\item $(v,v)$ is a positive element in $\mathcal{C}$,
\item $(v,v)=0\, \Leftrightarrow\,v=0$.
\end{enumerate}
\end{Def}

\indent An inner product $\mathcal{C}$-module $V$ has a norm defined as $||v||=\sqrt{||(v,v)||}$.  The norm completion of $V$ is called a \textit{Hilbert $\mathcal{C}$-module}.  

\begin{lem}\label{Hilbertmod}
The form $(\cdot, \cdot)$ on $\mathcal{E}_{<t/2}P$ satisfies:
\begin{enumerate}
\item $(F,Ee)=( F, E) e$ for $e\in P\mathcal{E}_0$,\
\item $\mathbb{C}$-linear in the right argument,\
\item $(v,w)^{*}=(w,v)$,\
\item $(v,v)$ is a positive element in $P\mathcal{E}_0$,\
\item and it's kernel is also an $\mathcal{E}_0$-bimodule and a $P\mathcal{E}_0$-right module.  
\end{enumerate}
\end{lem}
\begin{proof}  Checking $\mathbb{C}$ and right $\mathcal{E}_0$ linearity in the right argument is trivial when we notice $P\in\mathcal{E}_0'$.  Notice $(PE^{*}FP)^{*}=PF^{*}EP$.  It follows that property (3) above holds.  Property (4) follows easily.  From property (3), the left and right kernels coincide. To show the kernel is a $\mathcal{E}_0$-bimodule, we begin with assuming for all $F\in \mathcal{E}_{t/2}$ that $PF^{*}EP=0$.  Let $g,h\in \mathcal{E}_0$.  For all $F$, $(F,gEh)=PF^{*}gEhP=P(g^{*}F)^{*}EPh=(g^{*}F,E)h=0$.  This shows that the kernel is an $\mathcal{E}_0$-bimodule.  It being a right $P\mathcal{E}_{0}$-module is clear.
\end{proof}

\indent Since the kernel of $(\cdot,\cdot)$ is a $\mathcal{E}_0$-bimodule (right $P\mathcal{E}_0$-module), $\mathcal{E}_{<t/2}\,(\mathrm{mod}\, \mathrm{ker}\left(\cdot, \cdot)\right)\defeq \mathcal{E}$ is again a $\mathcal{E}_0$ bimodule (right $P\mathcal{E}_0$-module).  Letting $\mathcal{E}$ inherit $(\cdot,\cdot)$ as a $P\mathcal{E}_0$-valued sequilinear form, the closure of $\mathcal{E}$ (which we denote again as $\mathcal{E}$) is a Hilbert $P\mathcal{E}_0$-module.  

\indent We will denote the image of the code projection $P$ as $C\subset \mathcal{H}$.  Since $\mathcal{E}_0$ commutes with $P$, $C$ is a left $P\mathcal{E}_0$-module.  Thus we form $\mathcal{E}\otimes_{P\mathcal{E}_0} C$, and we can give it a sequilinear form defined on simple tensors by:
 \[\langle E\otimes v, F\otimes w\rangle= \langle  v, (E,F)w\rangle.\]
 
 \indent The form on the right in the above equation is the one on $\mathcal{H}$.  We then extend the form linearly.  After we show this form has trivial kernel, we complete $\mathcal{E}\otimes_{P\mathcal{E}_0} C$ with respect to the form (which we again denote as $\mathcal{E}\otimes_{P\mathcal{E}_0} C$).   
 
 \begin{lem} Suppose $\langle F\otimes w,  E\otimes v\rangle=0$ for all $F,w$.  Then $E\otimes v=0$.
 \end{lem}
 \begin{proof}  We recall a property of von Neumann algebras.  The projection onto the kernel of an operator in a von Neumann algebra is in the von Neumann algebra.   By weak operator closure, the projection onto the intersection of any number of kernels of operators in a von Neumann algebra is in the von Neumann algebra.  Thus we consider the intersection of all of the kernels of  $PF^{*}EP$ ranging over $F$.  Denote this closed subspace as $V$ and the projection onto it as $P_V$.  Notice $v\in V$.  Now $E\otimes v= E\otimes P_Vv=EP_V\otimes v$.  Notice that $EP_V=0$ in $\mathcal{E}$ since $(F, E) P_V=0$ for all $F$.
 \end{proof}
 
 \indent A corollary to the above lemma is that $(\cdot, \cdot)$ has trivial kernel.  We now prove an embedding theorem which is a generalization of the one given in the introduction.  
 
 \begin{thm} \label{codeembedding}The map $\mathcal{E}\otimes_{P\mathcal{E}_0} C\rightarrow \mathcal{E}_{<t/2}C$ defined by $E\otimes v\mapsto Ev$ is an isometry.  
 \end{thm}
 \begin{proof} We need only check that:
 \[\langle E\otimes v, F\otimes w\rangle= \langle v, (E,F) w\rangle= \langle v, PE^{*}FPw\rangle= \langle Ev,Fw\rangle.\]
 \end{proof}
 \indent This theorem also appears in \cite{KuWe}.  We make mention here that theorem \ref{firstembed} in the introduction is a special case of the above embedding theorem for $\mathcal{E}_0\simeq \mathbb{C}$.  
 \end{subsection}
 
 \begin{subsection}{Error Correction for Codes in Finite $W^*$-metric Spaces}
 \indent\indent We saw in the introduction that in the finite dimensional purely quantum case (i.e. $\mathcal{E}_0=\mathbb{C}I$), theorem \ref{codeembedding} (or just theorem \ref{firstembed}) was used to construct an error correcting transformation for code states supported on $P_C$.  We now generalize the notion of quantum error correction to codes in finite dimensional $W^*$-metric spaces to account for uncontrollable noise from a von Neumann algebra.  Let $\mathcal{E}_0$ be a von Neumann algebra in $\mathcal{B(H)}$.  Also, let $\mathcal{E}\subset\mathcal{B(H)}$ be an $\mathcal{E}_0$-bimodule containing $\mathcal{E}_0$.  Given a state $\rho\in\mathcal{B(H)}$, an error from $\mathcal{E}$ is any map of the form $\sum_i E_i\rho F^*_i$ where $E_i, F_i\in \mathcal{E}$.  Then we define an \textit{$\mathcal{E}_0$-error correcting transformation} for errors from $\mathcal{E}$ to be any CPTP map $\mathcal{R}$ on $\mathcal{B(H)}$ such that $\mathcal{R}(\sum_i E_i\rho F^*_i)=\sum_i e_i \rho f^*_i$ for some $e_i, f_i\in\mathcal{E}_0$ depending on the error from $\mathcal{E}$.  

\indent We again make the remark that we have already accepted that errors from $\mathcal{E}_0$ are inherently unprotectable.  Thus after $\mathcal{E}_0$ error correction we will only make measurements with random variables in $\mathcal{E}_0'$.  
 
\indent We now suppose that $P\in \mathcal{M}=\mathcal{E}_0'$ is a distance $t$ quantum code.  We recall that the quotient of $\mathcal{E}_{<t/2}$ with respect to the kernel of $(\cdot, \cdot)$, denoted $\mathcal{E}$, is a Hilbert $P\mathcal{E}_0$-module via the $P\mathcal{E}_0$-valued sequilinear form $(\cdot,\cdot)$.  We can give $\mathcal{E}$ a Hilbert space structure via the form $\langle E, F\rangle= Tr(E,F)$.  It follows from all the properties of $(\cdot, \cdot)$ from lemma \ref{Hilbertmod}  that $\langle \cdot, \cdot\rangle$ is a hermitian form.  We can see that for $e\in\mathcal{E}_0$ 
\begin{align*}
&\langle E, Fe\rangle= Tr(E, Fe)=Tr(PE^{*}FeP)=Tr(PE^{*}FPe)\\
&=Tr(PeE^{*}FP)=Tr(P(Ee^*)^{*}FP)=\langle Ee^*, F\rangle
\end{align*}

From the previous equations and the fact that $\mathcal{E}$ is a right representation of $P\mathcal{E}_0$, $\mathcal{E}$ is a $*$-representation of $P\mathcal{E}_0^{op}$.  To clarify, if $\mathcal{M}$ is a von Neumann algebra, $\mathcal{M}^{op}=\mathcal{M}$ as a complex vector space.  Yet multiplication in $\mathcal{M}^{op}$ is defined as $X\cdot_{\mathcal{M}^{op}}Y=YX$.  Thus right $\mathcal{M}$ modules are left $\mathcal{M}^{op}$ modules.  Since a finite dimensional von Neumann algebra $\mathcal{M}$ is isomorphic to $\mathcal{M}^{op}$ via the transpose operator, the finite dimensional representations of $\mathcal{M}^{op}$ are isomorphic to representations of $\mathcal{M}$.  Thus the following lemma follows from the classification of finite dimensional representations of finite dimensional von Neumann algebras in theorem \ref{finiteVN}.

\begin{lem} As a $P\mathcal{E}_0$-module, $\mathcal{E}\approxeq \bigoplus_i \mathcal{H}_i\otimes\mathcal{J}_i$ for finite dimensional Hilbert spaces $\mathcal{H}_i,\,\mathcal{J}_i$.  The action of $P\mathcal{E}_0$ is via the operators $\bigoplus_i \mathcal{B(H}_i)\otimes I_i$.  Here $I_i$ is the identity operator on $\mathcal{J}_i$.  
\end{lem}

Thus, we have the following corollary.

\begin{cor}\label{generatingset} There exists a $P\mathcal{E}_0$ generating set $\{E_\alpha\}$ for $\mathcal{E}$ such that $PE_i^*E_jP=PE_i^*E_iP\delta_{ij}$.  
\end{cor}
\begin{proof} We will choose any orthonormal basis $\{w^k_i\}\subset \mathcal{J}_i$.  For any non-zero $v_i\in \mathcal{H}_i$ $P\mathcal{E}_0(v_i\otimes w^k_i)=\mathcal{H}_i\otimes w^k_i$.  By construction, the subspaces $\mathcal{H}_i\otimes w^k_i$ for all $i,k$ are mutually orthogonal.  We choose $\{E_{ik}\}\subset \mathcal{E}$ to correspond to $v_i\otimes w^k_i$.  For simplicity, we will use a greek index instead of $ik$.  It follows that $\{E_\alpha\}$ forms a $P\mathcal{E}_0$ generating set for $\mathcal{E}$.  Also, it follows that $\langle E_\alpha e, E_\beta f\rangle=0$ if $\alpha\neq \beta$ for any $e,f \in P\mathcal{E}_0$.  In particular, $\langle E_\alpha, E_\beta f\rangle=Tr(E_\alpha,E_\beta f)=Tr((E_\alpha,E_\beta)f)=0$ if $\alpha\neq\beta$ for all $f\in \mathcal{E}_0$.  Since $Tr$ is a non-degenerate bilinear form on $P\mathcal{E}_0$, it follows that $(E_\alpha,E_\beta)=0$ for $\alpha\neq \beta$.  
\end{proof}

\indent We we will need another condition on this generating set, but first we will need a lemma regarding von Neumann algebras.  We mention that for von Neumann algebras, the spectral theorem holds for all self adjoint elements.  We only need this result here for finite dimensional elements, so we only make mention of that here.

\begin{thm}[\cite{OpAlg}] Given a self adjoint element $A$ of a finite dimensional von Neumann algebra $\mathcal{M}$, $A=\sum_i a_i P_i$ for $a_i\in\mathbb{R}\setminus 0$ and orthogonal projections $P_i\in\mathcal{M}$ (i.e. $P_iP_j=\delta_{ij} P_i$).
\end{thm}
\begin{lem}  Given a self-adjoint element $A=\sum_i a_i P_i$ in a von Neumann algebra $\mathcal{M}$, there exists an element $B\in\mathcal{M}$ such that $AB=BA=\sum_i P_i$.
\end{lem}
\begin{proof} Let $B= \sum_i \frac{1}{a_i} P_i$.
\end{proof}
\begin{lem}\label{generatingset2} There exist a $P\mathcal{E}_0$ generating set $\{E_\alpha\}$ for $\mathcal{E}$ such that $PE_\alpha^*E_\beta P=PE_\alpha^*E_\beta P\delta_{\alpha\beta}$, and $(E_\alpha,E_\beta)$ are projections in $\mathcal{E}_0$.  
\end{lem}
\begin{proof} We let $\{E_\alpha\}$ in the following come from corollary \ref{generatingset}.  Clearly $(E_\alpha,E_\alpha)$ is a positive element of $\mathcal{E}_0$, and thus $(E_\alpha,E_\alpha)=\sum_i a_i P_i$ for $a_i\in\mathbb{R}_{>0}$ and orthogonal projections $P_i$.  By the previous lemma, there exists a $B\in \mathcal{E}_0$ such that $B(E_\alpha,E_\alpha)=\sum_i P_i$.  Choosing $B$ as in the proof to the previous lemma, we see $B$ is a positive element.  Thus we define $\bar{E_\alpha}=E_\alpha\sqrt{B}$.  Here $\sqrt{B}=\sum_i\sqrt{\frac{1}{a_i}}P_i$.  Then we see 
\[(\bar{E_\alpha},\bar{E_\alpha})=P\sqrt{B}E_\alpha^*E_\alpha\sqrt{B}P=\sqrt{B}PE_\alpha^*E_\alpha P\sqrt{B}\]\[=\sqrt{B}(E_\alpha,E_\alpha)\sqrt{B}=B(E_\alpha,E_\alpha)=\sum_iP_i.\]
\end{proof}
\indent Given a code projector $P\in \mathcal{B(H)}$ supported on $C\subset \mathcal{H}$, we recall that we can form the tensor product of $P\mathcal{E}_0$-modules to form a Hilbert space $\mathcal{E}\otimes_{P\mathcal{E}_0}C$.    Notice that since $\mathcal{E}$ has a $P\mathcal{E}_0$ spanning set $\{E_\alpha\}$, we can write any element of $\mathcal{E}\otimes_{P\mathcal{E}_0}C$ as $\sum_\alpha E_\alpha\otimes v_\alpha$ for $v_\alpha\in C$.  Since $(E_\alpha,E_\beta)=0$ iff $\alpha\neq \beta$, the subspaces $E_\alpha\otimes C$ are all mutually orthogonal.  

\indent Also for each $E_i$, there exists a surjection $\mathcal{A}_\alpha: C\rightarrow E_\alpha\otimes C$ mapping $v\mapsto E_\alpha\otimes v$.  We can check that $\mathcal{A}_\alpha\mathcal{A}_\alpha^*$ is the projection onto $\mathcal{E}_\alpha\otimes C$.  Since $\langle E_\alpha\otimes v, A_\alpha w\rangle=\langle (E_\alpha,E_\alpha)v,w\rangle$, we can see  $\mathcal{A}_\alpha\mathcal{A}_\alpha^*(E_\alpha\otimes v)=\mathcal{A}_\alpha  (E_\alpha,E_\alpha)v=E_\alpha\otimes (E_\alpha,E_\alpha)v$.  Then since $\langle E_\alpha\otimes w, E_\alpha\otimes (E_\alpha,E_\alpha) v\rangle= \langle (E_\alpha,E_\alpha)w,(E_\alpha,E_\alpha)v\rangle=\langle (E_\alpha,E_\alpha)w,v\rangle=\langle E_\alpha\otimes w, E_\alpha\otimes v\rangle$ for all $w$, this implies $E_\alpha\otimes (E_\alpha,E_\alpha)v=E_\alpha\otimes v$.  Thus, $\sum_\alpha \mathcal{A}_\alpha\mathcal{A}_\alpha^*=I$ where $I$ is the identity on $\mathcal{E}\otimes_{\mathcal{E}_0} C$.  

\indent We now let $\mathcal{O}$ be any CPTP map from the orthogonal complement of $\mathcal{B}(\mathcal{E}\otimes_{P\mathcal{E}_0} C)$ in $\mathcal{B(H)}$ to $\mathcal{B}(C)$.  Now we have the following theorem.

 \begin{thm} \label{ect} The map $\mathcal{R}:\mathcal{B(H)}\rightarrow \mathcal{B}(C)$ defined by $\mathcal{R}(\rho)= \sum_\alpha \mathcal{A}^*_\alpha \rho \mathcal{A}_\alpha +\mathcal{O}(\rho)$ is an $\mathcal{E}_0$ error correcting transformation for any state $\rho$ supported on $C$.  
 \end{thm}
 \begin{proof}  We begin with any element of the form $|v\rangle\langle w|$ where $|v\rangle,|w\rangle \in C\subset\mathcal{H}$.  We consider any operators $E,F\in \mathcal{E}$, and then we express $E$ and $F$ in terms of elements of $\{E_\alpha\}$.  We will write this as $E=\sum_\alpha E_\alpha e_\alpha$ and $F=\sum_\alpha E_\alpha f_\alpha$ where $e_\alpha, f_\alpha \in \mathcal{E}_0$.  Now suppose there is an error on the code in the form of $E|v\rangle\langle w|F^*= \sum_{\alpha \beta} E_\alpha e_\alpha |v\rangle\langle w| f_\beta ^* E_\beta ^*$.  Now we consider 
 \begin{align}
 &\mathcal{R}(E|v\rangle\langle w|F^*)=\sum_\gamma \mathcal{A}^*_\gamma E|v\rangle\langle w|F^*\mathcal{A}_\gamma\\
 &= \sum_\gamma \mathcal{A}^*_\gamma\sum_{\alpha \beta} E_\alpha e_\alpha |v\rangle\langle w| f_\beta ^* E_\beta ^*\mathcal{A}_\gamma\\
 &= \sum_\gamma \mathcal{A}^*_\gamma E_\gamma e_\gamma |v\rangle \langle w| f^*_\gamma E^*_\gamma \mathcal{A}_\gamma\\
 &= \sum_\gamma (E_\gamma,E_\gamma)e_\gamma |v\rangle\langle w| f^*_\gamma(E_\gamma, E_\gamma).
 \end{align}
\indent Since $\sum_\gamma \mathcal{A}_\gamma \mathcal{A}^*_\gamma=I$, it follows that $\mathcal{R}$ is a CPTP transformation.

 \indent By the linearity of $\mathcal{R}$, the above argument is all that is needed to prove that $\mathcal{R}$ is an error correcting transformation for any error from $\mathcal{E}$ on any state supported on $C$.  
 \end{proof}

\end{subsection}  
\end{section}
\begin{section}{Conclusion and Discussion}
\indent\indent We recalled the definition of a $W^*$-filtration, $\{\mathcal{E}_t\}_{t\in\mathbb{R}\geq 0}$, on $\mathcal{B(H)}$ for a Hilbert space $\mathcal{H}$.  We also defined a $W^*$-metric on a von Neumann algebra $\mathcal{M}\subset\mathcal{B(H)}$ as a $W^*$-filtration with $\mathcal{M}'=\mathcal{E}_0$.  We also recalled the notion of a distance $t$ code in a $W^*$-metric space $\mathcal{M}$.  A distance $t$ quantum code $P\in\mathcal{B(H)}$, with support $C$, gave a quotient of the errors in $\mathcal{E}_{<t/2}$ the structure of a Hilbert $P\mathcal{E}_0$ module.  Denoting this quotient as $\mathcal{E}$, we then produced an isometric embedding theorem $\mathcal{E}\otimes_{P\mathcal{E}_0}C\hookrightarrow \mathcal{H}$.  This result generalizes the result in the finite dimensional case with $\mathcal{E}_0=\mathbb{C}I$. 

\indent We defined an error from $\mathcal{E}_{<t/2}$ on code states (i.e. states $\rho$ such that $\rho(P)=1$), and defined and produced and error correcting transformation $\mathcal{R}$ on such states.    We point out that the proof to this theorem is true for any operator supported on $C$.  With this in mind, we could take an alternate perspective to quantum error correction.  We could view errors from $\mathcal{E}_{<t/2}$ as happening on random variables in $\mathcal{M}$ that are supported on $C$.  In other words, random variables in $P\mathcal{M}P$.  In the case of studying finite dimensional $W^*$-metric spaces, making measurements with error corrected random variables in $P\mathcal{M}P$ is equivalent to making measurements with random variables in $P\mathcal{M}P$ on error corrected states supported on $C$.  Thus, we could define \textit{code random variables} as any element of $P\mathcal{M}P$, and then define a \textit{distance $<t/2$ error on code random variables} as any operator $\sum_i E_i A F_i^*$ for $E_i, F_i\in \mathcal{E}_{<t/2}$ and $A\in P\mathcal{M}P$.  Then an error correcting transformation is naturally defined as any CPTP map $\mathcal{R}:\mathcal{B(H)}\rightarrow P\mathcal{M}P$ such that $\mathcal{R}(\sum_i E_i A F_i^*)=\sum_i e_i A f^*_i=$ for $e_i,f_i\in \mathcal{E}_0$.  Of course the last equality above, by commutativity, is equal to $(\sum_{i}e_if^*_i)A$.  The same $P\mathcal{E}_0$ error correcting transformation in theorem \ref{ect} can be used here.  

\begin{subsection}{A Classical Example}

\indent\indent Since $W^*$-metrics on $\ell^{\infty}(M)$ for finite sets $M$ correspond to metrics $d$ on $M$, we end with pointing out how the error correcting transformation in theorem \ref{ect} correspond to a classical notion of error correction which we briefly recall.  Given a metric space $(M,d)$, a distance $t$ error on $M$ is simply any function $E: M\rightarrow M$ such that $d(E(x),x)\leq t$.  A distance $t$ code in $M$ is a subset $S$ where $d(x,y)\geq t$ for $x\neq y\in S$.  Thus a distance $t$ code $S$ satisfies $B_{t/2}(x)\cap B_{t/2}(y)=\emptyset$ for all $x\neq y\in S$.  Here $B_{t}(x)$ is the open ball of radius $t$ around $x$.  If $E$ is a distance $<t/2$ error, then we still have $E(x)\in B_{t/2}(x)$.  An error correcting transformation for $S$ is any function $R: M\rightarrow S$ satisfying $R(E(x))=x$ for all $x\in S$.  For example, any function $R$ such that $R(B_{t/2}(x))=x$ will work.  

\indent Consider a $W^*$-filtration $\{\mathcal{E}_t\}$ on $\mathcal{B}(\ell^2(M))$ where $\mathcal{E}_0=\ell^{\infty}(M)$.  Thus, as we denoted before, an element of $\ell^{\infty}(M)$ is written $\sum_{x\in M} a_x |x\rangle\langle x|$ for $a_x\in \mathbb{C}$.  Thus any projection in $\ell^{\infty}(M)$ can be written $P_S=\sum_{x\in S} |x\rangle\langle x|$ for some $S\subset M$.  We will point out that if $P_S$ is a distance $t$ code in $\ell^{\infty}(M)$, then $S$ is a set in $(M,d)$ whose distance between any two points is $\geq t$.  Since we know $\mathcal{E}_{<t}=\mathrm{span}\{|x\rangle\langle y| : \, d(x,y)< t\}$, the condition that $P_S \mathcal{E}_t P_S = \mathcal{E}_0 P$ implies $P_S \mathcal{E}_t P_S = \ell^{\infty}(S)$.  If there are $x,y\in S$ such that $d(x,y)< t$, then $P_S |x\rangle\langle y| P_S= |x\rangle\langle y|\notin \ell^{\infty}(S)$ which means $P_S$ is not a distance $t$ code.  

\indent Now we can easily check that for elements in $\mathcal{E}_{<t/2}$, $(|w\rangle \langle x|, |y\rangle\langle z|)=\delta_{wy}\delta_{xz} |x\rangle\langle x|$ if $x\in S$.  Thus, $\mathcal{E}$ is spanned by $\{|y\rangle\langle x|:\, x\in S,\, d(x,y)<t/2\}$.  Also, it is easy to see the right $P_S\ell^{\infty}(M)$ module generating set $\{E_\alpha\}$ from theorem \ref{generatingset2} is precisely this basis.  We will denote $|y\rangle\langle x|\in \mathcal{E}$ as $E_{yx}$.  Thus, we can see that $E_{yx}\otimes_{\ell^{\infty(S)}} S= \mathrm{span}(E_{yx}\otimes |x\rangle)$.  The map $\mathcal{A}_{yx}: S\rightarrow E_{yx}\otimes_{\ell^{\infty(S)} }S$ then maps $|w\rangle\mapsto E_{yx}\otimes |x\rangle\langle x||w\rangle$.  Thus, $\mathcal{A}_{yx}^*(E_{yx}\otimes w)=|x\rangle\langle x||w\rangle$ can be understood as moving $y\rightarrow x$ in $M$.  Thus, the entire error correcting transformation $\mathcal{R}$ can be understood as moving points in the open balls $B_{t/2}(x)$ to $x$ respectively for $x\in S$.

\end{subsection}

\end{section}

    \chapter[%
      Short Title of 3rd Ch.
   ]{%
     Code Construction in $\mathfrak{g}$-metric Spaces
   }%
   \label{ch:3rdChapterLabel}
   \indent\indent There is a class of  $W^*$-metrics coming from finite dimensional representations of complex Lie algebras.  Suppose $\pi:\mathfrak{g}\rightarrow\mathcal{B(H)}$ is a finite dimensional representation of $\mathfrak{g}$ where $\pi(\mathfrak{g})=\pi(\mathfrak{g})^{*}$.  We define a $W^*$-filtration $\{\mathcal{E}_t\}$ on $\mathcal{B(H)}$ as done for interaction algebras.  

\begin{Def}\label{g-metric}  Let $\pi:\mathfrak{g}\rightarrow\mathcal{B(H)}$, where $\pi(\mathfrak{g})=\pi(\mathfrak{g})^{*}$, be a representation of a $\mathfrak{g}$.  A \textit{$\mathfrak{g}$-metric} is a $W^*$-metric on $\mathbb{C}$ whose filtration terms are:
\begin{itemize}
\item $\mathcal{E}_0=\mathbb{C}I$\\
\item $\mathcal{E}_1=\pi(\mathfrak{g})$\\
\item $\mathcal{E}_t= \mathcal{E}_1^t$.
\end{itemize}
\end{Def}

We recall that $\mathcal{E}^t$ is defined as the linear span of the product of $\leq t$ operators from $\mathcal{E}$. We also note that for irreducible representations of $\mathfrak{g}$, there exists a $t$ such that $\mathcal{B(H)}=\mathcal{E}_t$.  This follows from Schur's lemma.

\indent In this chapter we will produce a code construction for $\mathfrak{su}(2)$-metric spaces coming from irreducible representations.  Although, the first part of the chapter we will be discussing a general outline for code constructions for \textit{multiplicity-free} representations of \textit{semi-simple} Lie algebras.    

\indent The construction we produce depends on three things.  First, we will need to find good classical codes in the \textit{weight diagram} of a representation.  Also, we will need a result about operators that are diagonal with respect to a \textit{weight basis} in a given representation.  Lastly, we will need Tverberg's theorem from convex geometry.

\indent Code constructions obviously give a lower bound of the maximal size (dimension) of a code.  In \cite{GNEC} a general code construction is given for (what we can interpret as) a finite dimensional $W^*$-metric space with the zero term $\mathcal{E}_0=\mathbb{C}I$.  We mention that (as one would hope) the code constructions for $\mathfrak{g}$-metric spaces given in this chapter give lower bounds to optimal codes that are better than the general one given in \cite{GNEC}.  Also, our codes for $\mathfrak{su}(2)$-metric spaces are constructive, unlike those in \cite{GNEC}.  Meaning, the construction will not rely on Tverberg's theorem, which is non-constructive in its proof.

\begin{section}{Review of Complex Semi-Simple Lie Algebras}
\indent\indent We will review a few results regarding representations of complex semi-simple lie algebras.  We'll refer the reader to  \cite{liealgebra} for example, for a more in depth discussion of Lie algebras.  

\begin{Def}  A \textit{Lie Algebra}, $\mathfrak{g}$, is a vector space with a bilinear operator $[\cdot,\cdot]:\mathfrak{g}\times \mathfrak{g}\rightarrow \mathfrak{g}$, called the \textit{Lie bracket} satisfying: 
\begin{itemize}
\item $[x,y]=-[y,x]$ (anti-symmetric),\
\item $[x,[y,z]]+[z,[x,y]]+[y,[z,x]]=0$ (Jacobi identity).
\end{itemize}
\end{Def}
Since we will only be considering Lie algebras over $\mathbb{C}$, we will assume this from here on.  

\indent The \textit{adjoint representation} of a Lie algebra $\mathfrak{g}$ is a representation $ad:\mathfrak{g}\rightarrow\mathcal{B}(\mathfrak{g})$ defined by $Ad[g](x)=[g,x]$.  The \textit{Killing form} of a Lie algebra is the bilinear form $K(x,y)=Tr(ad[x]ad[y])$.  There are many equivalent conditions for \textit{semi-simplicity} of a Lie algebra $\mathfrak{g}$, but the one we will state is that the Killing form on $\mathfrak{g}$ should be non-degenerate.  A \textit{Cartan subalgebra} $\mathfrak{h}$ of a Lie algebra $\mathfrak{g}$ is a nilpotent, self-normalizing subalgebra.  Meaning respectively: $ad[x]^n(y)=0$ for some $n$ and all $x,y\in\mathfrak{h}$; if $[g,x]\in \mathfrak{h}$ for all $x\in\mathfrak{h}$, then $g\in \mathfrak{h}$.  For semi-simple Lie algebras, a Cartan subalgebra $\mathfrak{h}$ is abelian (i.e. $[x,y]=0$ for $x,y\in\mathfrak{h}$), and it is diagonalizable with respect to the adjoint representation.   We can decompose a  semi-simple Lie algebra into its eigen spaces (called $\textit{root spaces}$) with respect to $ad[\mathfrak{h}]$: $\mathfrak{g}=\mathfrak{h}\oplus \bigoplus_{\alpha} \mathfrak{g}_\alpha$.  Here $\alpha\in\mathfrak{h}^*$ (the dual space of $\mathfrak{h}$) are called \textit{roots}, and they are defined by $ad[h](\mathfrak{g}_\alpha)=\alpha(h)\mathfrak{g}_\alpha$.  There is a nice structure theorem for finite dimensional representations of semi-simple Lie algebras.  

\begin{thm} Given a complex semi-simple Lie algebra $\mathfrak{g}$, let $\mathfrak{h}$ denote the Cartan subalgebra of $\mathfrak{g}$.  Let $V$ be a finite dimensional representation of $\mathfrak{g}$.  
\begin{enumerate}
\item $V$ can be decomposed into eigen spaces of $\mathfrak{h}$ called \textit{weight spaces}, $\bigoplus_{\lambda} V_{\lambda}$, for \text{weights} $\lambda\in\mathfrak{h}^*$ defined by $hV_\lambda=\lambda(h)V_\lambda$.\
\item If $g\in\mathfrak{g}$, $v\in V_{\lambda}$ and $gv\in V_{\gamma}\setminus\{0\}$, then $gv=\bar{g}v$ for a root vector $\bar{g}$ and $\gamma=\lambda+\alpha$ where $\alpha$ is the root corresponding to $\bar{g}$.  Also $gV_{\lambda}\subset V_{\gamma}$.\
\end{enumerate}
\end{thm}

\indent Suppose we have a finite dimensional module $V$ of a semi-simple Lie algebra $\mathfrak{g}$, where $V=\oplus_\gamma V_\gamma$ for weights $\gamma$ with corresponding weight spaces $V_{\gamma}$.  We construct a graph called the \textit{weight diagram} of $V$ by first letting the vertices be the weights.  If $v\in V_\lambda$, $g$ is a root vector and  $gv\in V_{\gamma}\setminus\{0\}$, then we connect $\lambda$ and $\gamma$ by an edge.  Now we endow this graph with the usual graph metric.  This means we assume edges have length equal to one, and then non-neighboring vertices have distance equal to the shortest edge path connecting them, and if there is no path the distance is infinite.    

\indent A complex associative algebra $\mathcal{A}$ is automatically a Lie algebra with bracket $[x,y]=xy-yx$.  If we are given a representation $\pi:\mathfrak{g}\rightarrow\mathcal{A}$ where $\pi[x,y]=[\pi(x),\pi(y)]$, then this induces an action of $\mathfrak{g}$ on $\mathcal{A}$ via $g(a)=[\pi(g),a]$ for $a\in\mathcal{A}$.  This action is a \textit{derivation}; by which we mean: $g(ab)=ag(b)+bg(a)$.  The \textit{universal enveloping algebra} of $\mathfrak{g}$, denoted $\mathcal{U}(\mathfrak{g})$, is an associative algebra where there exists a representation $i: \mathfrak{g}\rightarrow \mathcal{U}(\mathfrak{g})$ satisfying the following.  If $\pi:\mathfrak{g}\rightarrow\mathcal{A}$ is a representation in an associative algebra $\mathcal{A}$, then there exists a unique map $\pi':\mathcal{U}(\mathfrak{g})\rightarrow\mathcal{A}$ such that $\pi'(i(g))=\pi(g)$ for all $g\in \mathfrak{g}$.  We know that since $i$ is a representation of $\mathfrak{g}$ in the associative algebra $\mathcal{U}(\mathfrak{g})$, then $\mathfrak{g}$ has an action on $\mathcal{U}(\mathfrak{g})$.  We can understand how this action decomposes $\mathcal{U}(\mathfrak{g})$ into weight spaces if we can work with the following construction of $\mathcal{U}(\mathfrak{g})$.  We can identify $\mathcal{U}(\mathfrak{g})$ as a quotient of the tensor algebra of $\mathfrak{g}$, $\bigoplus_{n\geq 0}\bigotimes^n\mathfrak{g}$, by the relation $x\otimes y-y\otimes x=[x,y]$ for $x,y\in\mathfrak{g}$.  Since $\mathfrak{g}$ acts on $\mathcal{U}(\mathfrak{g})$ by derivations, we can notice $g(g_{\lambda_1}g_{\lambda_2}\ldots g_{\lambda_n})=(\lambda_1+ \lambda_2+\ldots \lambda_n)g_{\lambda_1}g_{\lambda_2}\ldots g_{\lambda_n}$ (where $g_\lambda$ is a root vector for root $\lambda$).  Thus the action of $\mathfrak{g}$ on the $\mathcal{U}(\mathfrak{g})$ can be decomposed into weight  spaces where all of the weights are sums of roots with multiplicity.  We will denote the weight space of $\lambda$ in $\mathcal{U}(\mathfrak{g})$ as $G(\lambda)$.  This decomposition is actually a $\mathbb{Z}^{\mathrm{rank}(\mathfrak{g})}$ grading.  By this we mean the product $G(\lambda_1)\cdot G(\lambda_2)\subset G(\lambda_1+\lambda_2)$.  We now define a length on the weights occurring in the action of $\mathfrak{g}$ on $\mathcal{U}(\mathfrak{g})$.  
\begin{Def} If $\lambda$ is weight occurring for the action of $\mathfrak{g}$ on $\mathcal{U}(\mathfrak{g})$, then we can express $\lambda=\sum_i \lambda_i$ ($\lambda_i$ being roots).  We define the length of $\lambda$, $\ell(\lambda)$, to be the minimal number of root vectors needed to form $\lambda$.  
\end{Def}

We now can state the following lemma.  

\begin{lem}\label{weightdiagram} Given an irreducible module $V=\oplus_\gamma V_\gamma$ of $\mathcal{U}(\mathfrak{g})$ where the sum is over all weights $\gamma$ occurring in the representation.  An operator $E_\lambda\in G(\lambda)$ will act on weight spaces $E_\lambda V_\gamma\subset V_{\lambda+\gamma}$.  Conversely, if an operator $E\in \mathcal{U}(\mathfrak{g})$ acts $EV_\gamma\subset V_{\gamma+\lambda}\, \forall \gamma$ then $E$ has a representative $\tilde{E}\in G(\lambda)$ through is action on $V$.
\end{lem}
\begin{proof}
Since for $h\in\mathfrak{h}$, $[h,E_\lambda]=hE_\lambda-E_\lambda h=\lambda(h)E_\lambda$, it is easy to check that $hE_\lambda V_\gamma=(\lambda+\gamma)(h)E_\lambda V_{\gamma}$.  Thus $E_\lambda V_\gamma \subset V_{\lambda+\gamma}$.  Now assume $EV_\gamma\subset V_{\gamma+\bar{\lambda}}\,\forall \gamma$, and $E=\sum_\lambda E_\lambda$ is a decomposition of $E$ with respect to the weight space decomposition of $\mathcal{U}(\mathfrak{g})$.  Now if $v_\gamma\in V_\gamma$ then $Ev_\gamma=\sum_\lambda E_\lambda v_\gamma$.  Since non-zero terms of the form $E_\lambda v_\gamma$ with respect to $\lambda$ are linearly independent (following from the first part of the paragraph), by assumption the only non-zero term has to be $E_{\bar{\lambda}} v_\gamma$.  Since this holds true for all $v_\gamma \in V_\gamma$ and all $\gamma$, we have $Ev=E_{\bar{\lambda}}v$ $ \forall v\in V$.  
\end{proof}

\begin{cor} An operator $E\in\mathcal{U}(\mathfrak{g})$ fixes all weight spaces of a representation $V$ iff there exists an operator $\tilde{E}\in G(0)$ such that $E=\tilde{E}$ when restricted to $V$.  
\end{cor}

\indent A \textit{multiplicity-free} representation of a semi-simple Lie algebra $\mathfrak{g}$ is a representation where all weight spaces are one-dimensional.  In multiplicity free, irreducible representations, $V$, $\mathcal{U}(\mathfrak{h})$ forms a maximal abelian sub-algebra of $\mathcal{B}(V)$.  Initially, we know it is an abelian $*$-subalgebra of $\mathcal{B}(V)$ since it is abelian in $\mathcal{U}(\mathfrak{g})$.  Maximality follows from the fact that $V$ decomposes into weight spaces that are all one-dimensional.  

\begin{subsection}{Irreducible Representations of $\mathfrak{su}(2)$}
\indent  The defining representation of the semi-simple Lie algebra $\mathfrak{su}(2)$ is on $\mathbb{C}^2$, and $\mathfrak{su}(2)$ is generated by operators: 
\begin{equation}
h=\begin{bmatrix}
\frac{1}{2} & 0\\
0                &-\frac{1}{2}
\end{bmatrix}\qquad
e=\begin{bmatrix}
0 & 1\\
0 & 0
\end{bmatrix}\qquad
f=\begin{bmatrix}
0 & 0\\
1 & 0
\end{bmatrix}
\end{equation}

The commutation relations are as follows:
\begin{equation}
[h,e]=e \qquad [h,f]=-f \qquad [e,f]=2h
\end{equation}

We make note that in many presentations the $h$ operator is replaced with $2h$.  We only use convention since in Chapter 4 it will make our presentation there simpler.  

\indent We now identity all irreducible representations of $\mathfrak{su}(2)$.  To give an identification the irreducible representations of $\mathfrak{su}(2)$, we first define the action of a general Lie algebra on the tensor product of two representations.  If $V$ and $W$ are two (finite dimensional) representations of a Lie algebra $\mathfrak{g}$, then $\mathfrak{g}$ has a representation on $V\otimes W$ via $g(v\otimes w)= gv\otimes w+ v\otimes gw$.  Then we extend the action linearly to all other tensors in $V\otimes W$.   

\indent Thus, inductively, $\mathfrak{su}(2)$ has a representation on the n-fold tensor product $\bigotimes^n\mathbb{C}^2$. It is easily checked that the symmetric tensors in $\bigotimes^n\mathbb{C}^2$, denoted $\bigvee^n \mathbb{C}^2$, form an invariant subspace with respect to the action of $\mathfrak{su}(2)$.  If $v=(1,0)\in\mathbb{C}^2$ (viewed as a row vector), then one can check that $v^n$ is a weight vector with weight $n/2$.  Also, $f^k(v^n)$ for $0\leq k\leq n$ forms a weight basis for $\bigvee^n \mathbb{C}$ and $e(v^n)=0$.  These, facts imply that $\bigvee^n \mathbb{C}^2$ is an irreducible representation of $\mathfrak{su}(2)$.  We denoted this representation as $V_{n/2}$.  We now state a theorem that these are all irreducible representations of $\mathfrak{su}(2)$.
\begin{thm}\label{stringrep}  \
\begin{enumerate}
\item The representations, $V_{n/2}$, of $\mathfrak{su}(2)$ are irreducible for all $n\geq 0$.  Each $V_{n/2}$ has a weight basis: 
\begin{equation}
v_\frac{n}{2},\, v_\frac{n-1}{2}, \ldots,\, v_\frac{n-k}{2},\, \ldots,\, v_\frac{-n}{2}
\end{equation}
where the subscript is the weight corresponding to the weight vector.  This basis is chosen such that  $f(v_{(n-k+1)/2})=(n-k)v_{(n-k)/2}$.  One can then check that $e(v_{(n-k)/2})=kv_{(n-k+1)/2}$. \\ 
\item The irreducible representations $V_{n/2}$ form all irreducible representations of $\mathfrak{su}(2)$.  Furthermore, all finite dimensional representations of $\mathfrak{su}(2)$ can be decomposed into the direct sum of irreducible representations.  
\end{enumerate}
\end{thm}
\end{subsection}
\end{section}

\begin{section}{Outline for Code Constructions}
\indent \indent The construction is a two part construction somewhat analogous to the general distance $t$ code construction given in \cite{GNEC}.  

\indent We begin with a $\mathfrak{g}$-metric space coming from a finite dimensional, multiplicity-free, irreducible representation of $\mathfrak{g}$.  We denote the $W^*$-filtration terms as $\mathcal{E}_\alpha$ for $\alpha\in \mathbb{R}_{\geq 0}$.  Our construction for a distance $t$ code will begin with finding a code projection $P$ that detects operators from $\left(\bigoplus_{\lambda\neq0}G(\lambda)\right)\cap \mathcal{E}_{<t}$.  We will pick a code projection $P$ that commutes with $G(0)$, which we know forms a maximal abelian sub-algebra diagonal with respect to the weight basis of $V$.  We have the following lemma. 

\begin{lem} Let $\mathcal{B}(V)$ be a $\mathfrak{g}$-metric space coming from a finite dimensional, multiplicity-free, irreducible representation.  A minimal distance $t$ set in the weight diagram for $V$ yields a distance $t$ code projection that commutes with $G(0)$ and detects operators from $\left(\bigoplus_{\lambda\neq0}G(\lambda)\right)\cap \mathcal{E}_{<t}$. 
\end{lem}

\begin{proof} Suppose we are given a code projection $P$ supported on weight vectors\\ $\{|v_1\rangle,|v_2\rangle,\ldots, |v_n\rangle\}$ that form a minimal distance $t$ set in the weight diagram.  We first note that $\mathcal{E}_{<t}$ and $G(\lambda)$ are both spanned by monomials in the root vectors of $\mathfrak{g}$.  Suppose $E\in G(\lambda)$ ($0<\lambda<t$) is a monomial in the root vectors.  Then $PEP=0$ since $E|v_i\rangle$ is proportional to a weight vector of distance $d$ away from $|v_i\rangle$ where $0<d<t$.  This follows from lemma \ref{weightdiagram}.  Thus $P$ detects operators in $\bigoplus_{0<\ell(\lambda)<t}G(\lambda)$.  Notice $\mathcal{E}_{<t}\subset \bigoplus_{\ell(\lambda)<t} G(\lambda)$.  Since $\mathcal{E}_{<t}$ and $G(\lambda)$ are spanned by monomials in the root vectors, we have $\left(\bigoplus_{\lambda\neq 0} G(\lambda)\right)\cap \mathcal{E}_{<t} = \bigoplus_{0<\ell(\lambda)<t} (G(\lambda)\cap \mathcal{E}_{<t})$.  Thus $P$ detects operators from $\left(\bigoplus_{0<\lambda < t} G(\lambda)\right)\cap \mathcal{E}_{<t}$.   
\end{proof}

\indent Now given $P$ from the previous lemma, we wish to find a code projection $P'\leq P$ that can detect operators from $G(0)\cap \mathcal{E}_{<t}$ as well.  This part of the construction is entirely analogous to the general code construction given in \cite{GNEC}.  We begin with a basis $\{E_1, E_2, \ldots, E_m\}$ of $G(0)\cap \mathcal{E}_{<t}$.  For simplicity sake we define $\overrightarrow{E}=(E_1, E_2, \ldots, E_m)$.  We let $\{|v_1\rangle,|v_2\rangle,\ldots, |v_n\rangle\}$ be the set of weight vectors supporting $P$.  Since $\{E_1, E_2, \ldots, E_m\}$ are all diagonal with respect to the weight basis, we have $\overrightarrow{E}|v_i\rangle=\overrightarrow{\alpha_i}|v_i\rangle$.  Here $\overrightarrow{\alpha_i}$ is the vector of eigen values of $\{E_1, E_2, \ldots, E_m\}$ for eigen vector $|v_i\rangle$.

\indent Now we wish to partition $\{1,2,\ldots, n\}$ into sets $X_j$ such that there exist unit vectors $|c_j\rangle=\sum_{k\in X_j} \beta_{kj}|v_k\rangle$ such that the following doesn't depend on $j$:
\[ \langle c_j|\overrightarrow{E}|c_j\rangle = \sum_{k\in X_j} |\beta_{kj}|^2\overrightarrow{\alpha_k}.\]

\indent Thus we let $P'$ be the projection supported on the $|c_j\rangle$, and the previous expression is equivalent to the error detection condition from operators in $G(0)\cap \mathcal{E}_{<t}$.

\indent The sum on the right hand side of the previous equation is a point in the convex hull of $\{\overrightarrow{\alpha_k}:k\in X_j\}$ for all $j$.  The following theorem of Tverberg gives conditions under when this is necessarily possible.  

\begin{thm}[Tverberg \cite{Tver}] \label{Tverberg} Given a set of $N$ points in $\mathbb{R}^n$, it is possible to partition the points into sets $\{P_1, P_2, \ldots, P_k\}$ where $conv(P_1)\cap conv(P_2)\cap \ldots \cap conv(P_k)\neq \varnothing$ provided $N\geq (n+1)(k-1)+1$.  
\end{thm}

\indent The above proof is a non-constructive one.  What we aim to do is actually produce a constructive partition that yields the result of the above theorem in the $\mathfrak{su}(2)$ case.  

\indent We end this section with a discussion about finding a basis for $G(0)\cap \mathcal{E}_{<t}$.  We will see for $\mathfrak{su}(2)$, $\{I, h, h^2, \ldots, h^{t-1}\}$ is such a basis.  We will see in the next section that all of the points $\overrightarrow{\alpha_k}\in\mathbb{R}^{n}$ will lay on the \textit{moment curve}.  We hypothesis for multiplicity-free representations of semi-simple Lie algebras: if $E\in G(0)\cap\mathcal{E}_ t$, then $E$ is equivalent to $\tilde{E}\in \mathcal{U}(\mathfrak{h})\cap \mathcal{E}_{t}$ when restricted to $V$.  This result is easy to prove for $\mathfrak{su}(2)$.  We don't necessarily need this result to show a code exists.  Although, after choosing a basis for $\mathfrak{h}$ (e.g. \textit{simple roots}), we can view this basis as a finite set of functions $\{h_i\}$ on weight vectors $\{|v_1\rangle,|v_2\rangle, \ldots |v_n\rangle\}$ (as above).  Then the entries of $\overrightarrow{\alpha_k}$ are determined by the products of $<t$ functions from $\{h_i\}$.  
\end{section}

\begin{section}{The $\mathfrak{su}(2)$ Case}

\indent\indent In this section we will implement the code construction from the previous section for irreducible representations of $\mathfrak{su}(2)$.  The construction will come from the following three observations.

\begin{thm} \label{thm1}For $\mathfrak{su}(2)$, $G(0)$ is spanned by $\mathcal{U}(\mathfrak{h})$ and the center $Z(\mathcal{U}(\mathfrak{su}(2)))$.  Furthermore, if $E\in G(0)$ is a polynomial in the standard generators of degree $<t$, then $E=Z+H$ for $Z\in Z(\mathcal{U}(\mathfrak{su}(2))),\,H\in \mathcal{U}(\mathfrak{h})$ and the degree of $Z$ and $H$ is $<t$.         
\end{thm}  

\begin{thm}\label{thm2} Arithmetic sequences form optimal minimal distance $d$ codes in the weight space of any irreducible representation of $\mathfrak{su}(2)$.
\end{thm}

\begin{thm}  \label{Tver} Suppose we have a set of $k$ points in $\mathbb{R}$ in arithmetic progression used to form vertices of a degree $d$ cyclic polytope.  The set of vertices can be partitioned into $s$ sets, $X_i$ for $ i=1,\ldots, s$, where $conv(X_1)\cap conv(X_2)\cap \ldots \cap conv(X_s)\neq \varnothing$ provided $k\geq (d+1)(s-1)$.  Furthermore, this can be done with a periodic $s$-coloring of the $k$ points, and there is an inductive formula for the convex coefficients of elements in $X_i$ for all $i$.  
\end{thm}

\indent Firstly, the theorem \ref{thm2} is actually very trivial but useful none-the-less.  For more complicated weight diagrams, finding good minimal distance $t$ sets becomes a much harder problem.  Secondly, the theorem \ref{thm1} is not even true for any other semi-simple lie algebras, but that is why we discussed using weight multiplicity free representations in the previous section.  We'll next give a proof of a the first theorem above. 

\begin{proof} (\textit{Proof of theorem \ref{thm1}}) Given $E\in G(0)$, $E$ can be written as a non-commutative polynomial in the standard generators of $\mathfrak{su}(2)$, namely $e,f$ and $h$.  We will show that for $\mathfrak{su}(2)$, $G(0)=\mathcal{U}(\mathfrak{h})\cdot Z(\mathcal{U}(\mathfrak{su}(2)))$ inductively on the degree of $E$.  By the Poincare-Birkoff-Witt theorem $E$ can be written as a linear combination of $h^\alpha e^\beta f^\gamma$.  Since the summands written in this form are known to be linearly independent and they are also eigen vectors of $h$ in the adjoint action of $\mathfrak{su}(2)$ on $\mathcal{U}(\mathfrak{su}(2))$, it must be that $[h,h^\alpha e^\beta f^\gamma]=0$.  Thus all summands of $E$ must be of the form $h^\alpha e^\beta f^\beta$.  Now a term $h^\alpha e^\beta f^\beta$ can be written as $h^\alpha (ef)^\beta$ up to terms of order $<\alpha+2\beta$ by using the bracket relation $[e,f]=ef-fe=2h$.  The casimir operator $C=h^2+\frac{1}{2}(ef+fe)$ is the known generator of $Z(\mathcal{U}(\mathfrak{su}(2)))$.  Using the lie relations it is easy to see $ef=C-h^2+\frac{1}{2}h\in \mathcal{U}(\mathfrak{h})\cdot Z(\mathcal{U}(\mathfrak{su}(2)))$.  
\end{proof}

\indent Given an irreducible representation $V_\lambda$ of $\mathfrak{su}(2)$ of highest weight $\lambda$, we use theorem \ref{thm2} to form a distance $d$ code in the weight space of $V_\lambda$.  By theorem \ref{thm1}, we now have to find a code based on the corresponding weight vectors that detects operators from the set $h,h^2,\ldots, h^{d-1}$.  In the next section we will introduce cyclic polytopes and prove theorem \ref{Tver}. 

\begin{subsection}{Tverberg Point for Cyclic Polytopes Formed by Arithmetic Sequences}
\indent\indent We begin by introducing cyclic polytopes.  The \textit{moment curve} in $\mathbb{R}^d$ is the image of the function $m: \mathbb{R}\rightarrow \mathbb{R}^d$ defined by $m_d(t)=(t,t^2,\ldots, t^d)$.  A cyclic polytope is the convex hull of a set $\{m_d(t):\,t\in T\subset \mathbb{R}\}$ for any given subset $T\subset\mathbb{R}$.   Following theorem \ref{thm2}, we are interested in when the elements of $T$ form an arithmetic sequence. We will denote the cyclic polytope with vertices $\{m_d(t):\,t\in T\subset{R}\}\subset \mathbb{R}^d$ as $C_d(T)$.

\indent We introduce some terminology.  Suppose we are given points $S\subset\mathbb{R}^d$, and  there exists a partition of $S$ into $s$ sets, $S_1, S_2,\ldots, S_s$, such that the convex hull of each set $S_j$, denoted $\mathrm{conv}(S_j)$, satisfies $\bigcap_{j=1}^{s}\mathrm{conv}(S_j)\neq \O$.  We call a partition of this type a \textit{Tverberg partition for (S,s)}, and a point $p\in\bigcap_{j=1}^{s}\mathrm{conv}(S_j)$ a \textit{Tverberg point} for (S,s). 

\indent In proving theorem \ref{Tver} we will actually find a Tverberg point for the presented partition.  We begin proving this theorem through a series of easy lemmas.  The first lemma says that a Tverberg partition for points on a moment curve is invariant under affine transformations of $\mathbb{R}$.

\begin{lem}\label{affineindependent}
Let $T\subset \mathbb{R}$ and assume there exists a partition, $\{T_i; i=1,2,\ldots,s\}$, of $T$ so that there is an $x\in \bigcap_{i=1}^s \mathrm{conv}(m_d(T_i))$.  Then for any affine transformation of $\mathbb{R}$, $f(x)=ax+b$, there exists an $x'\in \bigcap_{i=1}^s \mathit{conv}(m_d(f(T_i)))$.  Moreover, the convex coefficients of elements of $m_d(T_i)$ yielding $x$ are also unchanged for elements of $m_d(f(T_i))$ yielding $x'$.  
\end{lem}

\begin{proof}  The proof is straightforward.  Since $\{m_d(T_1),m_d(T_2),\ldots, m_d(T_s)\}$ is a Tverberg partition of $m_d(T)$ then a Tverberg point for this partition satisfies a set of equations
\begin{equation}
\sum_{x\in T_j}\alpha_x x^i=\sum_{x\in T_k}\alpha_x x^i\quad \forall j,k\; \mathrm{and}\, 1\leq i \leq d.
\end{equation}
Now consider an affine transformation $f(x)=ax+b$ of $\mathbb{R}$.  For $\vec{x}\in\mathbb{R}^d$, denote the $i$-th coordinate of $\vec{x}$ as $x_i$ and notice
\begin{align*}
&\left[\sum_{x\in T_j}\alpha_x m_d(f(x))\right]_i = \sum_{x\in T_j}\alpha_x (ax+b)^i=\sum_{x\in T_j}\alpha_x \sum_{r=0}^{i}\binom{i}{r}a^rb^{i-r}x^{r}\\ 
&=\sum_{r=0}^{i}\binom{i}{r}a^rb^{i-r}\sum_{x\in T_j}\alpha_x x^{r}=\sum_{r=0}^{i}\binom{i}{r}a^rb^{i-r}\sum_{x\in T_k}\alpha_x x^{r}=\sum_{x\in T_k}\alpha_x (ax+b)^i\\
							&=\left[\sum_{x\in T_k}\alpha_x m_d(f(x))\right]_i.
							\end{align*}
Since these equalities hold for any $j,k$,  the lemma is proved.	
\end{proof}							
\indent We now know from this lemma that finding a Tverberg point for $m_d(T)$ where $T$ forms an arithmetic sequence is no harder than finding a Tverberg point for $m_d(T')$ when $T'=\{0,1,2,\ldots,\#(T)-1\}$.  Thus the following sections will prove theorem \ref{Tver} for this case.  

\begin{subsubsection}{Inductive Step}

\indent \indent Our plan for this section is to show how to construct a Tverberg point, $\vec{v}$, when partitioning $m_d(\{0,1,\ldots, (N-1)+(S-1)\})$ into $S$ sets when we know how to construct a Tverberg point for $m_{d-1}(\{0,1,\ldots, N-1\})$ when partitioning into $S$ sets.  The partitions will come from an alternating $S$-coloring of $\mathbb{Z}$ through arithmetic sequences. \\
\indent Before continuing, it will be convenient to make a few notational definitions.  To ease the onset of variable suffocation in this section, we will assume $d$ and $S$ are fixed.  Now we notationally define:
\begin{align*}
\mathbb{N}_N &\stackrel{\text{\tiny def}}{=}\{0,1,2,\ldots,N-1\}\quad \forall N\in \mathbb{N},\\
A_j&\stackrel{\text{\tiny def}}{=} A\cap \{j+Sn\,|\,n\in\mathbb{Z}\}\,\mathrm{for}\,A\subset\mathbb{R}\,\mathrm{and}\,j\in \mathbb{R},\\
\vec{B}  &\stackrel{\text{\tiny def}}{=} \left(1,\binom{d}{2},\binom{d}{3},\ldots, \binom{d}{d-1}\right)\in\mathbb{R}^{d-1},\\
\vec{v}_k (\in\mathbb{R}^{d-1})&\stackrel{\text{\tiny def}}{=}\mathrm{Tverberg\, point\, for}\, (m_{d-1}(\mathbb{N}_N+k),S),\\
\alpha^j_k(t)&\stackrel{\text{\tiny def}}{=} \mathrm{coefficient\,found\,in}\quad\vec{v}_k=\sum_{t\in\{\mathbb{N}_N+k\}_j}\alpha^j_k(t)m_{d-1}(t),\\
\overrightarrow{(k,j)}&\stackrel{\text{\tiny def}}{=}\sum_{t\in \{\mathbb{N}_{N}+k\}_j} \alpha^j_k(t)m_d(t)
\end{align*}

\indent In the above definitions $j$ should be understood as the partition color, and $k$ can be understood as a variable that slides a set of points along the moment curve.

\indent The argument to find $\vec{v}_k$ and $\alpha^j_k(t)$ will use induction on $d$, and so $\vec{v}_k$ will be properly defined later.  Although, by lemma \ref{affineindependent} we know how to find $\vec{v}_k$ and $\alpha^j_k(t)$ if we can find $\vec{v}_0$ and $\alpha^j_0(t)$. Lemma \ref{affineindependent} gives the following identity for the coefficients $\alpha^j_k(t)$.
\begin{equation}
\alpha^j_0(t)=\alpha^{j+1}_{1}(t+1)=\ldots=\alpha^{j+S-1}_{S-1}(t+S-1).
 \end{equation}  

The idea to find the Tverberg point $\vec{v}\in \mathbb{R}^d$ is fairly simple.  If $P:\mathbb{R}^d\rightarrow \mathbb{R}^{d-1}$ the projection onto the first $d-1$ coordinates, then clearly $P(m_d(T))=m_{d-1}(T)$.  Notice that by inductive assumption the points in $P(\mathrm{conv}\{\overrightarrow{(k,j)}\,|\,0\leq j,k\leq S-1\})$ are all Tverberg points for $m_{d-1}(\mathbb{N}_{N+S-1})$ for the alternating partition described in theorem \ref{Tver}.  Seeing this, we seek non-negative numbers, $\{c(k)\}_{0\leq k \leq S-1}$, where $\sum_k c(k)=1$, and they satisfy  
\begin{equation}\sum_{0\leq k\leq S-1}c(k)\overrightarrow{(k,j)}\quad= \sum_{0\leq k\leq S-1}c(k)\overrightarrow{(k,l)}\end{equation}
 for all partition colors $0\leq j,l\leq S-1$.  We know for any such $\{c(k)\}_{0\leq k \leq S-1}$, the first $d-1$ coordinates of the vectors in the previous equation will match.  Thus we only have to find coefficients such that the $d-th$ coordinates match.  In other words we need \begin{equation}\label{dthcoord}\sum_{0\leq k\leq S-1}c(k)\left[\overrightarrow{(k,j)}\right]_d\quad= \sum_{0\leq k\leq S-1}c(k)\left[\overrightarrow{(k,l)}\right]_d.\end{equation}
For simplicity sake, we denote $[\overrightarrow{(k,j)}]_d\stackrel{\text{\tiny def}}{=}(k,j)$.  Another way to write equation \ref{dthcoord} is 
\begin{equation}
\mathbf{M}=
\begin{bmatrix}
(0,0) & (1,0) & (2,0) & \ldots & (S-1,0)\\
(0,1) & (1,1) & (2,1) & \ldots & (S-1,1)\\
\vdots & \vdots & \vdots & \ddots & \vdots \\
(0,S-1) & (1,S-1) & (2,S-1) & \ldots & (S-1,S-1)
\end{bmatrix}
 \;\vec{\mathbf{c}}\,=\, r\,
\begin{bmatrix}
1\\
1\\
\vdots\\
1
\end{bmatrix}.
\end{equation}

Here $r$ is a some real number and $\vec{\mathbf{c}}\in \Delta_{S-1}$, the standard $(S-1)$-dimensional simplex in $\mathbb{R}^S$.
Now a lemma that simplifies the above matrix.

\begin{lem}
\begin{equation}
(k,j)=\vec{B}\cdot \vec{v}_{k-1}+(k-1,(j-1)\mod S)
\end{equation}
where $\vec{B}\cdot \vec{v}_{k-1}$ is the standard dot product.
\end{lem}
\begin{proof}  We begin with the expression for $(k,j)$
\begin{align*}
 &=\sum_{t\in \{\mathbb{N}_{N}+k\}_j} \alpha^j_k(t)t^d\\
&= \sum_{t\in\{\mathbb{N}_{N}+k\}_{j-1}}\alpha^{j}_k(t+1)(t+1)^d + \left\{\alpha^j_k(k)k^d \, \mathrm{if} \,k\in\mathbb{Z}_j\right\}\\
\end{align*}
If $N-1+k\in\mathbb{Z}_{j-1}$ this term isn't considered in the previous expression.
\begin{align*}
&=\sum_{t\in\{\mathbb{N}_{N}+k\}_{j-1}}\alpha^{j-1}_{k-1}(t)\left[ \sum_{m=0}^{d-1}\binom{d}{m}t^m+t^d\right]\\
&\qquad\qquad+\left\{\alpha^{j-1}_{k-1}(k-1)\left[ \sum_{m=0}^{d-1}\binom{d}{m}(k-1)^m+(k-1)^d\right] \,\mathrm{if}\,k\in\mathbb{Z}_j\right\}\\
&=\sum_{t\in\{\mathbb{N}_{N}+k-1\}_{j-1}}\alpha^{j-1}_{k-1}(t)\left[ \sum_{m=0}^{d-1}\binom{d}{m}t^m+t^d\right]\\
&=\sum_{m=0}^{d-1}\binom{d}{m}\sum_{t\in\{\mathbb{N}_{N}+k-1\}_{j-1}}\alpha^{j-1}_{k-1}(t)t^m+\sum_{t\in\{\mathbb{N}_{N}+k-1\}_{j-1}}\alpha^{j-1}_{k-1}(t)t^d\\
&=\sum_{m=0}^{d-1}\binom{d}{m}\left[\vec{v}_{k-1}(d-1)\right]_m+(k-1,(j-1)\mod S)\\
&=\vec{B}\cdot \vec{v}_{k-1}+(k-1,(j-1)\mod S)
\end{align*}
\end{proof}

We can now use the previous lemma iteratively to show that the matrix $\mathbf{M}$ can be written in the following form.  
\begin{equation}
\begin{bmatrix}
(0,0)& (0,s-1)+ \vec{B}\cdot\vec{v}_0& \ldots & (0,1)+\vec{B}\cdot\left[\vec{v}_0+\ldots+\vec{v}_{s-2}\right]\\
(0,1) & (0,0)+ \vec{B}\cdot\vec{v}_0& \ldots & (0,2)+\vec{B}\cdot\left[\vec{v}_0+\ldots+\vec{v}_{s-2}\right]\\
(0,2)& (0,1)+ \vec{B}\cdot\vec{v}_0&\ldots & (0,3)+\vec{B}\cdot\left[\vec{v}_0+\ldots+\vec{v}_{s-2}\right]\\
\vdots & \vdots &  \ddots & \vdots \\
(0,s-2) & (0,s-3)+ \vec{B}\cdot\vec{v}_0&\ldots &(0,s-1)+\vec{B}\cdot\left[\vec{v}_0+\ldots+\vec{v}_{s-2}\right]\\
(0,s-1) & (0,s-2)+ \vec{B}\cdot\vec{v}_0 & \ldots & (0,0)+\vec{B}\cdot\left[\vec{v}_0+\ldots+\vec{v}_{s-2}\right]
\end{bmatrix}
\end{equation}

\indent Notice now that $\mathbf{M}$ has the form of a circulant matrix modified by adding different constants to each column.  Thus the row sums are all the same.  Now it is clear that we can let $c(k)=\frac{1}{s}\;\forall k$, and equation \ref{dthcoord} will be satisfied.  

\indent At this point we only need to find an $N$ such that $\mathbb{N}_N$ has an alternating partition into $S$ sets whose convex hulls have a common intersection point.  It is easy to see that, for the alternating partition, $2S-1$ points are needed in order for the partition sets to have a common point in their convex hulls.  

\begin{cor} The points $m_d(\mathbb{N}_{(d+1)(S-1)})$ can be partitioned via the alternating partition into $S$ sets whose convex hulls all have a common intersection point.  The number of points needed to satisfy such criteria on the moment curve matches the theoretical upper bound for the number of points needed given by Tverberg's Theorem. 
\end{cor} 
\end{subsubsection}
\end{subsection}
\end{section}

\begin{section}{Conclusions and Discussion}
\indent\indent We found a constructive method to find a Tverberg point for a particular set of points on the moment curve in $\mathbb{R}^d$.  In turn, this helped up produce a distance $t$ code of dimension $k$ in a $\mathfrak{su}(2)$-metric space with Hilbert space of dimension $N$, provided $N> (t+1)^2(k-1)$.  We didn't beat Tverberg's non-constructive number of points needed to have a partition into $k$ sets in $\mathbb{R}^d$, but we matched it.  We conjecture that the $N$ needed to partition $m_d(\{0,1,\ldots,N-1\})$ into $k$ sets whose convex hulls have nontrivial intersection isn't going to be less than the Tverberg upper bound $(d+1)(k-1)+1$.  This is only a crude assumption from the complexity of cyclic polytopes.  For example, they maximize the number of facets a polytope can have with a fixed number of vertices.  

\indent  In \cite{GNEC}, there was a construction that stated if the dimension of the error space is $D$, then there exists a code of dimension $k$ provided the dimension of the code space was $>kD(D+1)$.  Since the dimension of $\mathcal{E}_t$ (ignoring the identity) is $(t+1)^2$, our construction produces a code of dimension $k$ provided the code space has dimension $>(k-1)D$.  So again, at least we beat the general lower bound on the dimension of the code space needed to produce a dimension $k$ code that can detect an error space of dimension $D$.

\end{section}

    \chapter[%
      Short Title of 4th Ch.
   ]{%
     MacWilliams Type Identities for $\mathfrak{su}$-metric Spaces
   }%
   \label{ch:4thChapterLabel}

\indent\indent Given a finite dimensional $W^*$-metric space $\mathcal{M}$, the problem of finding sharp upper bounds for the dimension of a distance $t$ code is generally a hard one.  For commutative $\mathcal{M}$, the problem is equivalent to finding upper bounds for the size of minimal distance $t$ sets in a finite metric space.  Perhaps the only upper bound for the size of a distance $t$ code in a generic metric space is given by a volume bound.  For example, if we are given a finite metric space $(M,d)$ where all balls of radius $<t/2$ are isometric (e.g. a lattice on a torus), then any distance $t$ code $C$ must satisfy $|C|\leq |M|/|B_{<t/2}|$ ($\mathit{|X|}$ is the volume of $X\subset M$ with the counting measure).  Here $|B_{<t/2}|$ is the number of points in the ball of radius $<t/2$.  Theorem \ref{codeembedding} is a generalization of such volume bound.  In the purely quantum case (i.e. $\mathcal{M}$ is $\mathcal{B(H)}$), theorem \ref{codeembedding} for distance $t$ codes, $C$, where $\mathcal{E}=\mathcal{E}_{<t/2}$ implies $\dim(C)\leq \dim(H)/\dim(\mathcal{E}_{t/2})$.  

\indent On the other hand, there are powerful techniques for finding upper bounds on the size of  distance $t$ linear codes in $\mathbb{F}_2^n$.   The techniques involve setting up a linear programming problem using the MacWilliam's identity \cite{ECC}.  To give a flavor of such techniques, we'll give brief overview of the MacWilliams identities and the linear programming problem.  We first introduce the distance distribution for a subset of a finite metric space.
\begin{Def}  Given a finite metric space $(M,d)$ and a subset $C\subset M$, the \textit{distance distribution} of the set $C$ is given by:
\[ B_t(C)=\frac{1}{|C|}\# \{(x,y)\in C\times C; d(x,y)=t\}\]
\end{Def}

\indent The MacWilliams identity relates the distance distributions of a linear code $C\subset \mathbb{F}_2^n$ to that of its dual code $C^{\perp}$.  We introduce the \textit{enumerator polynomial} for $C\subset \mathbb{F}_2^n$: 
\[B(C;x,y)=\sum_{i=0}^n B_i(C) x^iy^{n-i}.\]
The MacWilliams identity is:
\[B(C^{\perp};x,y)=\frac{1}{|C|}B(C;(x+y),(x-y)).\]

The linear programming problem we set up to find an upper bound on the size of a distance $t$ code in $\mathbb{F}_2^n$ begins with maximizing  $\sum_i B_i(C)=|C|$.  The constraints come from the MacWilliams identity (using the fact that $B_i(C^{\perp})\geq 0$ for all $i$), and the following:
\begin{itemize}
\item $B_0(C)=1$,\\
\item $B_i(C)=0$ for $1\leq i< t$,\\
\item $B_i(C)\geq 0$ for $i\geq t$.
\end{itemize}

The second set of constraints above come from the requirement that $C$ be a distance $t$ code.  

\indent When the $W^*$-filtration is the quantum Hamming metric, quantum analogues of the  MacWilliams identities have been already used to produce linear programming problems to find upper bounds for quantum codes (see \cite{Rains02} and \cite{QMac}).  In this chapter we will extend such techniques for finding upper bounds to codes in $\mathfrak{su}(2)$-metric spaces.  

\begin{section}{Quantum Distance Distribution}
\indent\indent We wish to define a \textit{quantum distance distribution}.  The quantum distance distribution is a straight forward generalization of one of the weights defined in \cite{QMac}.  We will show that in the classical case of finite metric spaces, the quantum distance distribution is a generalization of the classical distance distribution. 

\indent We begin by defining a class of Hermitian forms on a matrix algebra $B(\mathbb{C}^n)$.  Fixing operators $X,Y\in B(\mathbb{C}^n)$, we define:
\begin{equation}B_{XY}(a,b)\defeq Tr(a^*XbY).\end{equation}

\indent Consider a $W^*$-metric on $B(\mathbb{C}^n)$ with filtration terms $\{\mathcal{E}_t\}$.  Recall that the \textit{pure} $d$-th filtration term is defined as $\mathcal{E}_d/\mathcal{E}_{<d}$.  Since $B(\mathbb{C}^n)$ is a Hilbert space with Hilbert-Schmidt (HS) form being the inner product, we can canonically map the $d-th$ filtration term into $B(\mathbb{C}^n)$.  Now we have an orthogonal direct sum decomposition: 
\begin{equation} \label{nothing}
B(\mathbb{C}^n)=\bigoplus_{d\in\mathbb{R}_{\geq 0}} \mathcal{E}_d/\mathcal{E}_{<d}.
\end{equation}

We can now define a quantum distance distribution.
\begin{Def}\label{quantumdd}  The \textit{quantum distance distribution} of distance $d$ with respect to the $W^*$-filtration $\{\mathcal{E}_t\}$ is the bilinear form:
\begin{equation}B_{d}(X,Y)=Tr_{d}B_{XY}(\cdot, \cdot).\end{equation}
Here the notation $Tr_d$ denotes taking the trace of $B_{XY}(\cdot, \cdot)$ over the pure $d$-th filtration term with respect to the HS form.
\end{Def}

\indent The distance distribution $B_{d}(X,Y)$ up to normalization corresponds precisely $B_d(X,Y)$ in equation [4] of \cite{QMac}.  There $B_d(X,Y)$ is called a \textit{quantum weight} in analogy to distance distributions for linear codes in $\mathbb{F}_2^n$ being called weight distributions.  In \cite{QMac}, the quantum weight is:
\[B_d(X,Y)=\frac{1}{Tr(XY)}\sum_{E_d}Tr(E^*_dXE_dY).\]
The sum is over all multi-Pauli matrices, $\{E_d\}$, with exactly $d$ tensor terms not equal to the identity, and after an overall scaling  factor this is an orthonormal basis for the pure $d$-th filtration term in the quantum Hamming metric.

\begin{subsection}{Example:  The Classical Case}
\indent \indent We begin with a $W^*$-metric coming from a classical finite metric space $(M,d)$.  We will consider a representation of $\ell^{\infty}(M)$ in $\mathbb{B}(\ell^2(M))$.  Here it is somewhat pointless to make a distinction in notation between $\ell^2(M)$ and $\ell^{\infty}(M)$, but we will do so just to distinguish between the vector space and multiplication operators on the vector space. The $W^*$-filtration terms we denote as $\ell^{\infty}(M)=\mathcal{E}_0\subset \mathcal{E}_{\alpha_1}\subset \ldots\subset \mathcal{E}_{\alpha_k}=\mathcal{B}(\ell^2(M))$.  Given a subset $S\subset M$, we consider this as a projection $P_S\in \mathcal{E}_0\subset B(\ell^2(M))$.  The projection has the form: 
\[P_S=\sum_{x\in S}|x\rangle\langle x|.\]
\indent Now we consider an orthonormal basis of $\mathcal{E}_{\alpha_j}/\mathcal{E}_{\alpha_{j-1}}$ of the form $\{|x\rangle\langle y|: d(x,y)=\alpha_j\}$.  Thus: 
\begin{align}
B_{\alpha_j}(P_S,P_S)&=\sum_{d(x,y)=\alpha_j}Tr(|x\rangle\langle y| P_S |y\rangle\langle x|P_S)\\
&=\sum_{d(x,y)=\alpha_j}\langle x|P_S|x\rangle\langle y|P_S|y\rangle\\
&=\sum_{d(x,y)=\alpha_j; x,y\in S} 1\\
&=\#\{(x,y)\in S\times S: d(x,y)=\alpha_j\}
\end{align}

\indent Thus the quantum distance distribution evaluated at $(P_S, P_S)$ for a subset $S$ of a finite metric space is proportional to the classical distance distance distribution of $S$.  

\end{subsection}

\end{section}
\begin{section}{Quantum Weights}
\indent\indent In \cite{QMac} two quantum weights were introduced.  One quantum weight we have already shown to be a generalization of a distance distribution for classical metric spaces.  We understood the quantum distance distributions as the trace of a sesquilinear form over the pure $W^*$-filtration terms of a finite dimensional $W^*$-metric space.  We now introduce another sesquilinear form on $\mathcal{B(H)}$ (for finite dimensional $\mathcal{H}$) in variables $a,b$: 
\[A_{XY}(a,b)=Tr(Xa^*)Tr(Yb).\]

We now define a weight $A_d(X,Y)$ that is a generalization for arbitrary finite dimensional $W^*$-metric spaces of the weight given in equation 3 of \cite{QMac}.  

\begin{equation}A_d(X,Y)=Tr_d A_{XY}(\cdot, \cdot).\end{equation}

\indent In \cite{Rains02}, the following theorem is shown.
\begin{thm}\label{weightrelation} Given a projection $P$ and an operator $M$ in $M_n(\mathbb{C})$ we have the following. If the image of $P$ has dimension $K$ then 
\[K\mathrm{Tr}(M^* P M P)\geq |\mathrm{Tr}(MP)|^2\geq 0.\]
\end{thm} 

\indent The essence of the proof in \cite{Rains02} is the following.  If we let $v$ be a normalized, uniformly random vector from the image of $P$, then the expectation value 
\[E(|\langle v|M|v\rangle|-\frac{1}{K}\mathrm{Tr}(PM)|^2)\geq 0.\]
It is then shown that:
\[E(|\langle v|M|v\rangle|-\frac{1}{K}\mathrm{Tr}(PM)|^2)=\frac{1}{K(K+1)} \left(K\mathrm{Tr}(M^* PMP)-|\mathrm{Tr}(MP)|^2\right).\]

\indent If we are given a $W^*$-metric space with trivial zero term, then the condition for a code to detect an operator $M$ is $PMP=\alpha(M) P$ where $\alpha(M)\in \mathbb{C}$.  Since $\alpha(M)=\dim(P)^{-1}Tr(PM)$, we can see theorem \ref{weightrelation} implies $K\mathrm{Tr}(M^* P M P)= |\mathrm{Tr}(MP)|^2$ iff $P$ detects $M$.  If $P$ is a distance $d$ code for such a $W^*$-metric space then we have: 
\begin{equation}Tr(P)B_k(P,P)=A_k(P,P)\end{equation}
for $W^*$-filtration terms $k< d$.

\indent We make mention that if the zero term is generated by more operators besides the identity, then the previous equality of quantum weights isn't necessarily true.  For instance, if we have a $W^*$-metric space whose zero term is $\ell^{\infty}(M)$ that comes from a classical metric space $(M,d)$ and code projection $P\in \ell^{\infty}(M)$ then $A_t(P,P)=0$ for all $t>0$.  Meanwhile, as we previously showed, $B_t(P,P)=\#\{(x,y): d(x,y)=t\}$.

\indent The goal of the remainder of this chapter is to give a linear relationship between the weights $\{A_d\}$ and $\{B_d\}$ for $\mathfrak{su}(2)$-metric spaces coming from irreducible representation of $\mathfrak{su}(2)$.  Then, we will use theorem \ref{weightrelation} to set up a linear programming problem to find an upper bound for the dimension a fixed distance code in a given $\mathfrak{su}(2)$-metric space.  

\end{section}
\begin{section}{Preliminaries}
\begin{subsection}{Clebsch-Gordan Coefficients and Wigner $6j$-Symbols}
\indent\indent We will review the \textit{Clebsch-Gordan decomposition} of the tensor product of two irreducible representations of $\mathfrak{su}(2)$ and the \textit{Wigner $6j$ symbols}.  We only will state results needed, but the information here on Clebsch-Gordan coefficients and Wigner $6j$-symbols are easily accessible in the literature.  We give \cite{QTAM} or \cite{6jCnQ} as possibilities. 

\indent  In chapter 3, theorem \ref{stringrep} displayed all irreducible representations of highest weight $2j$, $V_j$, of $\mathfrak{su}(2)$.  All these representations can be identified as symmetric tensor powers of the defining representation $V_{1/2}=\mathbb{C}^2$.  As $\mathbb{C}^2$ is a Hilbert space, so are all symmetric powers of $V_{1/2}$.  It is easily checked that the weight basis for $V_j$ in theorem \ref{stringrep} is an orthogonal basis, but yet it isn't  normalized.  We use the bra and ket notation to specify an orthonormal weight basis of $V_j$.  A normalized weight vector in $V_j$ of weight $m$ is denoted $|jm\rangle$.  This basis is precisely specified by the \textit{Condon Shortley phase convention}.  First the highest weight vector $|jj\rangle=v_{1/2}^j$ is specified. Then we apply the lowering operator $f$ to recursively define $f|jm\rangle=C_-(jm)|j(m-1)\rangle$ (we make mention that $e|jm\rangle = C_+(jm)|j(m+1)\rangle$).  The phase convention is to take $C_\pm(a,j)=\sqrt{a(a+1)-j(j\pm1)}$. 

\indent The Clebsch-Gordan decomposition supplies a $\mathfrak{su}(2)$-module isometric embedding $V_j\hookrightarrow V_a\otimes V_b$ (where $j= |a-b|, |a-b|+1,\ldots, |a+b|$) that is unique up to a phase.   A vector of weight $m$ in the $V_j$ subrepresentation of $V_a\otimes V_b$ is denoted $|(ab)jm\rangle$, and it's choice of phase is given in:  

\[|(ab)jm\rangle = \sum_{k=\min(m-b,-a)}^{a} C^{(ab)jm}_{k} |ak\rangle\otimes|b(m-k)\rangle.\]

The coefficients $C^{(ab)jm}_k$, succinctly denoted $C^{jm}_k$, are the Clebsch-Gordan coefficients.  The Condon Shortley phase convention specifies that $C^{jj}_a \in \mathbb{R}_{>0}$.  Then, all other coefficients $\{C^{jm}_k\}$ are specified by the requirement that $e|(ab)jj\rangle=0$ (so as to be a highest weight vector), the normalization condition, and the application of the lowering operator.  

\indent  Next, we will consider two orthonormal bases of the $\mathfrak{su}(2)$ invariant subspace of 
$V_a\otimes V_b \otimes V_c \otimes V_d$ (if it exists) and a unitary map between the two.  We will let $|(ab)e(cd)e\rangle$ denote the $\mathfrak{su}(2)$-invariant tensor found by identifying $V_e$ as a subrepresentation of $V_a\otimes V_b$ and $V_c\otimes V_d$ (assuming triangle inequalities are satisfied).  Now after tensoring these two copies of $V_e$, we can find a single $\mathfrak{su}(2)$-invariant normalized vector.  The coefficients of this vector with respect to simple tensors of the weight bases for $V_a,\,V_b,\,V_c$ and $V_d$ have phases satisfying the Condon Shortley convention. Likewise, we let $|(ad)f(bc)f\rangle$ be the invariant tensor found by first identifying $V_f$ as a subrepresentation of $V_a\otimes V_d$ and $V_b\otimes V_c$.  Then we find the invariant tensor as a subrepresentation of $V_f\otimes V_f$.  The two $\mathfrak{su}(2)$ invariant basis of $V_a\otimes V_b\otimes V_c\otimes V_d$ are $\{|(ab)e(cd)e\rangle:\,e\}$ and $|(ad)f(bc)f\rangle:\,f\}$ (we consider all $e$ and $f$ satisfying appropriate triangle inequalities).  The unitary map between the two bases is given by the following relations:
\[|(ab)f(cd)f\rangle=\sum_{e} (-1)^{a+b+c+d}\sqrt{(2e+1)(2f+1)}\left\{ \begin{matrix} a&b& f\\ c &d& e \end{matrix}\right\}|(ad)e(bc)e\rangle.\]
 The quantities specified by the curly brackets are the \textit{Wigner $6j$ symbols}. 
 
 \indent Two more facts will be necessary to understand.  One, the $\mathfrak{su}(2)$ module isometry $\rho_a: V_{a}\rightarrow V^{*}_{a}$ we will be using is:
 \[\rho_a(|a(a-j)\rangle)=(-1)^{j} \langle a(j-a)|.\]   
 It is easy to check that this is indeed a $\mathfrak{su}(2)$ module isometry.  The map $\rho_{1/2}$ maps the defining representation of $\mathfrak{su}(2)$ to its dual. Then $\rho_a$ is the $2a$-th tensor power of $\rho_{1/2}$ when viewing $V_a$ as the $2a$-th symmetric tensor power of the defining representation.  
  
\indent The second, if we take the tensor $|(\ell_1\ell_2)dj\rangle$ and switch the two tensor terms to identify another element, denoted as $|(\ell_2\ell_1)dj\rangle$, in $V_{\ell_1}\otimes V_{\ell_2}$ (this makes sense when $\ell_1=\ell_2$), then $|(\ell_2\ell_1)dj\rangle=(-1)^{2\ell-d}|(\ell_1\ell_2)dj\rangle$.  This follows easily from observing the Clebsch-Gordan coefficients in $|(\ell_1\ell_2)dj\rangle$.

\end{subsection}

\begin{subsection}{Basis for Pure $\mathfrak{su}(2)$-filtration Terms}
\indent \indent We begin with a $2\ell+1$ dimensional irreducible module, $V_\ell$,  of $\mathfrak{su}(2)$ where $\ell \in \{0,\frac{1}{2}, 1, \frac{3}{2}, \ldots\}$.  This representation can be used to define another $\mathfrak{su}(2)$-module, $\mathcal{B}(V_\ell)$, defined via the adjoint action by an element of $\mathfrak{su}(2)$. Meaning, if $g\in\mathfrak{su}(2)$ and $M\in B(V_\ell)$, then $g\cdot M = [g,M]$.  Since $V_\ell \otimes V^{*}_\ell\simeq \mathcal{B}(V_\ell)$ and $V^*_{\ell}\simeq V_{\ell}$ as $\mathfrak{su}(2)$-modules, the representation on $\mathcal{B}(V_\ell)$ is isomorphic to the representations $V_\ell \otimes V _\ell \approxeq V_0 \oplus V_1 \oplus V_2 \oplus \ldots \oplus V_{2\ell}$.  The last isomorphism is the \textit{Clebsch-Gordan decomposition}.  

\begin{lem} The subspace $\mathcal{E}_d$ from the $W^*$-metric on $B(V_\ell)$ associated to the $\mathfrak{su}(2)$ representation is $V_0\oplus V_1 \oplus \ldots \oplus V_d$ understood as seen through the Clebsch-Gordan isomorphism.  Thus $\mathcal{E}_d/\mathcal{E}_{d-1}=V_d$. 
\end{lem}
\begin{proof}  Since $\mathfrak{su}(2)$ acts on $\mathcal{B}(V_{\ell})$ by derivations by the lie bracket on $\mathcal{B}(V_{\ell})$, it is easy to check that polynomials of degree $\leq d$ in $e,f$ and $h$ form an invariant subspace of $\mathcal{B}(V_{\ell})$.  It is also clear that $\{I, e, e^2, \ldots, e^d\}$ are all highest weight vectors (i.e. $[e,e^k]=0$).  In the following we use the \textit{adjoint operator} defined as $ad[f](x)=[f,x]$. We can check that $ad^{2k+1}[f](e^k)=0$, $ad^{2k}[f](e^k)\neq 0$, and $ad^{j}[f](e^k)$ $\forall j$ and $0\leq k \leq d$ spans polynomials of degree $\leq d$ in $e,f$ and $h$.  It follows from the classification of finite dimensional irreducible representations of $\mathfrak{su}(2)$ that $e^k$ is a highest weight vector for a sub-representation of $\mathcal{B}(V_{\ell})$ isomorphic to $V_{k/2}$.  Thus, the result follows.  
\end{proof}

\indent We denote the orthonormal weight basis of $V_d$ (as a subspace of $B(V_\ell)$) as \\ $\{M_d,M_{d-1},\ldots, M_{-d}\}$.  Here $M_k$ is the weight vector of weight $2k$.  From the previous lemma we see that $M_k\propto ad^{2(d-k)}[f](e^d)$, and we will give the proportionality constants later in the chapter.  For the sake of some simplification of formulas, we will re-scale the quantum distance distributions $B_d$ and quantum weights $A_d$ for the $\mathfrak{su}(2)$-metric space $\mathcal{B}(V_\ell)$.  
\[A_d(X,Y)= \frac{1}{\sqrt{2d+1}} \sum_{i=-d}^{d} Tr(M^{*}_iX) Tr(M_iY).\]
\[B_d(X,Y)= \frac{1}{\sqrt{2d+1}}\sum_{i=-d}^{d} Tr(M^{*}_iXM_iY).\]

\begin{rem} These weight operators are defined for all $0\leq d\leq 2\ell$.  We will also be referring to elements of $V_d$ as being elements in $B(V_\ell)$, and should be understood as the image of $V_d$ as seen through the Clebsch-Gordan isomorphism.  When a weight vector $M_i$ is referred to, it will be understood from context from which $V_d$ it came from.  
\end{rem}

\indent In $B(V_\ell)$, $\frac{e^d}{||e^d||}$ (the norm $||\cdot||$ is from the HS hermitian form) is a normalized highest weight $d$ vector.  Thus the image of $|dd\rangle$ through $V_d\hookrightarrow V_\ell\otimes V_\ell \simeq B(V_\ell)$ is, up to a phase factor $\alpha$, $\frac{e^d}{||e^d||}$.  Since the matrix coefficients of $e^d$ with respect to the weight basis of $V_\ell$ are real and non-negative, we know that $\alpha=\pm 1$.  From the expressions for $A_d$ and $B_d$, we see that this sign factor won't make a difference to our calculations, so we can let: 
\[M_d=\frac{e^d}{||e^d||}\]
and recurively
\[M_{d-j-1}=\frac{1}{C_-(d,d-j)}ad[f](M_{d-j}).\]  

\end{subsection}

\end{section}
\pagebreak
\begin{section}{Main Theorem}
 \begin{thm} 
\begin{enumerate}\indent

\item The bilinear forms $A_d$ and $B_d$ are $SU(2)$ (in turn $\mathfrak{su}(2)$) invariant.  Equivalently, they can be viewed as $SU(2)$ invariant maps \[B(V_\ell)\otimes B(V_\ell)\rightarrow \mathbb{C}.\] 
Since $(B(V_\ell)\otimes B(V_\ell))^*\approxeq V_\ell\otimes V_\ell\otimes V_\ell\otimes V_\ell$, they can be viewed as $\mathfrak{su}(2)$ invariant tensors in $V_\ell\otimes V_\ell\otimes V_\ell\otimes V_\ell$.\
\item \[A_d=\sum_{e} (-1)^{2\ell-e}\sqrt{(2e+1)(2d+1)}\left\{ \begin{matrix} \ell&\ell& d\\ \ell &\ell& e \end{matrix}\right\}B_e.\]
 
\end{enumerate}

\end{thm}

\begin{proof}  Let us first just consider a bilinear form $\mathcal{B}$ on $B(V_\ell)$.  By saying it is $SU(2)$-invariant we mean $\mathcal{B}(gXg^{-1}, gYg^{-1})=\mathcal{B}(X,Y)$ $\forall X,Y$.  This is clearly true for $B_d$ and $A_d$.  For example, for $g\in SU(2)$:
\begin{align*}
B_d(gXg^{-1},gYg^{-1})&=\sum_{i=-d}^{d} Tr(M^{*}_igXg^{-1}M_igXg^{-1})\\
&=\sum_{i=-d}^{d} Tr((g^{-1}M_ig)^{*}X(g^{-1}M_ig)Y)\\
&=\sum_{i=-d}^{d} Tr(M^{*}_iXM_iY).
\end{align*}
The last equality comes from the previous comment that this quantity could be understood as the trace of a sesquilinear form over $V_d$, and this representation of $SU(2)$ is a unitary one with respect to the Hilbert-Schmidt form.  

\indent Now an invariant bilinear form $\mathcal{B}$ can be understood as a $SU(2)$ invariant element in $\left(B(V_\ell)\otimes B(V_\ell)\right)^{*}$.  In general, $SU(2)$ invariant elements in $B(V_\ell)^{*}\otimes B(V_\ell)^{*}$ can also be understood as $SU(2)$-module homomorphisms $B(V_\ell)\rightarrow B(V_\ell)^{*}\approxeq B(V_\ell)\approxeq V_\ell\otimes V_\ell \approxeq V_0\oplus V_1 \oplus \ldots \oplus V_{2\ell}$.  It follows from Shur's lemma that any $SU(2)$-module homomorphism $B(V_\ell)\rightarrow B(V_\ell)$ must decompose as the block sum of operators between isomorphic, irreducible components of $B(V_\ell)$.  This means if $M\in V_m$ and $N\in V_n$ where $n\neq m$, then  $\mathcal{B}(M,N)=0$.  Thus $\mathcal{B}$ decomposes as the sum of bilinear forms on irreducible components of $B(V_\ell)$.  

\indent Now again if $M\in V_m$ and $N\in V_n$ where $n\neq m$, then $Tr(M^{*}N)=0$ since $V_m$ and $V_n$ are orthogonal subspaces of $B(V_\ell)$.  Thus the bilinear operator $A_d$ is only non-zero on $V_d$.  Now consider the $\mathfrak{su}(2)$ invariant tensor $|(\ell_1 \ell_2^{*})d(\ell_3\ell_4^{*})d\rangle$ whose phase is chosen such that $I\otimes \rho\otimes I\otimes \rho(|(\ell_1 \ell_2^{*})d(\ell_3\ell_4^{*})d\rangle)=|(\ell_1\ell_2)d(\ell_3\ell_4)d\rangle$.  Here the numerical subscripts to $\ell$ specify tensor order in $B(V_\ell)\otimes B(V_\ell)$.  Since $A_d$ is a $\mathfrak{su}(2)$ invariant bilinear form on $V_d\subset B(V_\ell)$, it can be identified as a multiple, $a_d$, of the dual tensor to $|(\ell_1 \ell_2^{*})d(\ell_3\ell_4^{*})d\rangle$.

\indent We now notice that:
 \begin{align}
 &Tr(M^{*}|x_1\rangle\langle x_2|M|x_3\rangle\langle x_4|)=\langle x_4|M^{*}|x_1\rangle\langle x_2|M|x_3\rangle\\
 &= Tr(M^{*}|x_1\rangle\langle x_4|)Tr(M|x_3\rangle\langle x_2|),
 \end{align}
 so $B_d(|x_1\rangle\langle x_2|, |x_3\rangle\langle x_4|)= A_d(|x_1\rangle\langle x_4|,|x_3\rangle\langle x_2|)$.  This relation between $A_d$ and $B_d$ tells us that $B_d$ can be identified as a multiple, $b_d$, of the dual tensor to $|(\ell_1\ell^{*}_{4})d(\ell_3\ell^{*}_{2})d\rangle$, again, whose phase is chosen such that $I\otimes \rho\otimes I\otimes \rho(|(\ell_1\ell^{*}_{4})d(\ell_3\ell^{*}_{2})d\rangle)=|(\ell_1\ell_{4})d(\ell_3\ell_2)d\rangle$.  What we will say at this moment is $a_d=b_d\defeq \alpha_d$.  The reason being is that if we switch the second and fourth tensor terms in $|(\ell_1\ell_2)d(\ell_3\ell_4)d\rangle$ we get $|(\ell_1\ell_4)d(\ell_3\ell_2)d\rangle$.  This follows the how the left/right ordering of the tensor terms in $|(\ell_1\ell_2)d(\ell_3\ell_4)d\rangle$ and $|(\ell_1\ell_4)d(\ell_3\ell_2)d\rangle$ are considered in the Clebsh-Gordon decomposition. \\
\indent We take notice now that $|(\ell_1\ell_4)d(\ell_3\ell_2)d\rangle=(-1)^{2\ell-d}|(\ell_1\ell_4)d(\ell_2\ell_3)d\rangle$.  Thus, we can now state:
\[\frac{A_d}{\bar{\alpha_d}}=\sum_{e} (-1)^{2\ell-e}\sqrt{(2e+1)(2d+1)}\left\{ \begin{matrix} \ell&\ell& d\\ \ell &\ell& e \end{matrix}\right\}\frac{B_e}{\bar{\alpha_e}}.\]
We make comment that the $6j$ symbols are real numbers, and thus appear without conjugation.

\indent We now calculate $\bar{\alpha}=A_d(|(\ell_1\ell^{*}_2)d(\ell_3\ell_4^{*})d\rangle)$.  First, we find a matrix expression for $|(\ell_1\ell^{*}_2)d(\ell_3\ell_4^{*})d\rangle=I\otimes\rho^{-1}\otimes I \otimes \rho^{-1}(|(\ell_1\ell_2)d(\ell_3\ell_4)d\rangle)$.  Notice that:
\[ |(\ell \ell)dd\rangle = \sum_{k=-\ell}^{\ell} C^{(\ell\ell) dd}_{k} |\ell k \rangle |\ell (d-k)\rangle\xrightarrow{I\otimes \rho}\sum_{k=-\ell}^{\ell} (-1)^{\ell-d+k}C^{(\ell\ell) dd}_k |\ell k \rangle\langle \ell (k-d)|.\]

\indent Also notice that:
\[|(dd)0\rangle = \sum_{k=-d}^{d}C^{dd0}_k |d k\rangle|d(-k)\rangle.\]
We can compute the Clebsh-Gordon coefficients to be $C^{dd0}_k=\frac{(-1)^{d-k}}{\sqrt{2d+1}}$. Thus we can see, 
 \[I\otimes \rho\otimes I\otimes \rho (|(\ell_1\ell_2)d(\ell_3\ell_4)d\rangle)=\sum_{k=-d}^{d}\frac{(-1)^{d-k}}{\sqrt{2d+1}}M_k\otimes M_{-k}.\]
 Thus, 
 \begin{align*}
 \bar{\alpha}&=A_d(|(\ell_1\ell_2^{*})d(\ell_3\ell_4^{*})d\rangle)\\
 &= \frac{1}{\sqrt{2d+1}}\sum_{k,j=-d}^{d} \frac{(-1)^{d-k}}{\sqrt{2d+1}}Tr(M^{*}_jM_k)Tr(M_j M_{-k})\\
&=\frac{1}{\sqrt{2d+1}} \sum_{k=-d}^d \frac{(-1)^{d-k}}{\sqrt{2d+1}} Tr(M_k M_{-k})\\
 &= \frac{1}{\sqrt{2d+1}} \sum_{k=-d}^d \frac{(-1)^{d-k}}{\sqrt{2d+1}}\left(\prod_{\substack{d-k+1\leq m\leq d \\ -d+k+1\leq n\leq d}} C_-(d,m)C_-(d,n)\right)^{-1}  \cdot \\
 &\hspace{5cm} Tr\left(ad^{k}[f](\frac{e^d}{||e^d||}) ad^{2d-k}[f](\frac{e^d}{||e^d||} )\right)\\
 &=\frac{1}{\sqrt{2d+1}}\sum_{k=-d}^{d} \frac{(-1)^{d}}{||e^d||^2\sqrt{2d+1}}\left(\prod_{\substack{d-k\leq m\leq d \\ -d+k\leq n\leq d}} C_-(d,m)C_-(d,n)\right)^{-1}Tr(e^dad^{2d}[f](e^d))
 \end{align*}
 The last line follows from $Tr$ being proportional to the Killing form.
Now it is a simple calculation to see:
\[Tr(e^{d}ad^{2d}(e^{d}))=(-1)^{d}\prod_{-d+1\leq j\leq d} C_-(d,j)Tr(e^{d}f^{d})\]
and
\[ ||e^d||^2=Tr(f^de^d).\]
We also can compute,
\begin{align*}
&\prod_{\substack{d-k+1\leq m\leq d \\ -d+k+1\leq n\leq d}} C_-(d,m)C_-(d,n)\\
&=\left(\prod_{\substack{d-k+1\leq m\leq d \\ -d+k+1\leq n\leq d}} (d+m)(d-m+1)(d+n)(d-n+1)\right)^{1/2}\\
&=(2d)!,
\end{align*}
and
\begin{align*}
&\prod_{-d+1\leq j\leq d} C_-(d,j)\\
&=\left(\prod_{-d+1\leq j\leq d} (d+j)(d-j+1)\right)^{1/2}\\
&=(2d)!.
\end{align*}
Thus we have, $\alpha_d=1$.  This yields our desired result. 
\end{proof}
\end{section}

\begin{section}{Conclusions and Discussion}
\indent\indent  We found a linear relationship between weight enumerators for $\{A_m\}$ and $\{B_n\}$ for $\mathfrak{su}(2)$-metrics on $\mathcal{B(H)}$.  This involved using the well known Wigner 6j-symbols.  This relationship in turn can be used to set up the following linear programming problem to find the upper bound on the dimension of a distance $t$ code in $B(V_{\ell})$.  We give an example for $\ell=5/2$ and $t=2$.  We wish to find the largest $k\leq 2\ell$ such that for $(B_0,B_1,B_2,B_3,B_4,B_5)\in \mathbb{R}^6_{\geq 0}$:
\begin{align*}
&kB_0=-\sum_{e} (-1)^{e}\sqrt{(2e+1)}\left\{ \begin{matrix} 5/2&5/2& 0\\ 5/2&5/2& e \end{matrix}\right\}B_e\\
&kB_1=-\sum_{e} (-1)^{e}\sqrt{3(2e+1)}\left\{ \begin{matrix} 5/2&5/2& 1\\ 5/2&5/2& e \end{matrix}\right\}B_e\\
&kB_2=-\sum_{e} (-1)^{e}\sqrt{5(2e+1)}\left\{ \begin{matrix} 5/2&5/2& 2\\ 5/2&5/2& e \end{matrix}\right\}B_e\\
&kB_3\geq -\sum_{e} (-1)^{e}\sqrt{7(2e+1)}\left\{ \begin{matrix} 5/2&5/2& 3\\ 5/2&5/2& e \end{matrix}\right\}B_e\\
&kB_4\geq -\sum_{e} (-1)^{e}\sqrt{9(2e+1)}\left\{ \begin{matrix} 5/2&5/2& 4\\ 5/2&5/2& e \end{matrix}\right\}B_e\\
&kB_5\geq -\sum_{e} (-1)^{e}\sqrt{11(2e+1)}\left\{ \begin{matrix} 5/2&5/2& 5\\ 5/2&5/2& e \end{matrix}\right\}B_e.\\
\end{align*}

\indent To solve such a problem, we would begin by fixing a $k\in\{1,2,3,4,5\}$.  Computer programs, such as SAGE, can be used to compute the Wigner $6j$-symbols and find a solution to this linear programming problem.

\end{section}


       
   \backmatter
   
   \bibliographystyle{halpha}
   \bibliography{DissertationBibliography}

\begin{thebibliography}{RHSS97}

\bibitem[Bla06]{OpAlg}
Bruce Blackadar.
\newblock {\em Operator Algebras}, volume 122 of {\em Operator Algebras and
  Non-Commutative Geometry III}.
\newblock Springer-Verlag, 2006.

\bibitem[CFS95]{6jCnQ}
J.~Scott Carter, Daniel~E. Flath, and Masahico Saito.
\newblock {\em The Classical and Quantum 6j Symbols}.
\newblock Mathematical Notes. Princeton University Press, 1995.

\bibitem[Cho75]{choi}
M.~Choi.
\newblock Completely positive linear maps on complex matrices.
\newblock {\em Linear Algebra and Its Applications}, pages 285--290, 1975.

\bibitem[Con95]{spectraltriple}
Alain Connes.
\newblock Geometry from the spectral point of view.
\newblock {\em Letters in Mathematical Physics}, 34:203--238, 1995.

\bibitem[CRSS97]{QECorthgeo}
A.R. Calderbank, E.M. Rains, P.W. Shor, and N.J.A. Sloane.
\newblock Quantum error correction and orthogonal geometry.
\newblock 78:405--409, 1997.

\bibitem[CRSS98]{GF(4)codes}
A.R. Calderbank, E.M. Rains, P.W. Shor, and N.J.A. Sloane.
\newblock Quantum error correction via codes over gf(4).
\newblock {\em IEEE}, 44(4):1369--1387, July 1998.

\bibitem[CS91]{SPLG}
J.H. Conway and N.J.A. Sloane.
\newblock {\em Sphere Packing, Lattices and Groups}, volume 290 of {\em A
  Series of Comprehensive Studies in Mathematics}.
\newblock Springer-Verlag, 1991.

\bibitem[CS96]{CodesExist}
A.R. Calderbank and P.W. Shor.
\newblock Good quantum error-correcting codes exist.
\newblock {\em Phy. Rev. A}, (54):1098--1105, 1996.

\bibitem[Jac79]{liealgebra}
Nathen Jacobson.
\newblock {\em Lie Algebras}.
\newblock Dover Publications, Inc., New York, 1979.

\bibitem[Kan42]{Kant1}
L.V. Kantorovic.
\newblock On the translation of masses.
\newblock {\em C. R. (Doklady) Acad. Sci. URSS (N.S.)}, 37:199--201, 1942.

\bibitem[KL96]{QECC}
Emanuel Knill and Raymond Laflamme.
\newblock A theory of quantum error-correcting codes.
\newblock Apr 1996, arXiv:quant-ph/9604034v1.

\bibitem[KLV99]{GNEC}
Emanuel Knill, Raymond Laflamme, and Lorenza Viola.
\newblock Theory of quantum error correction for general noise.
\newblock Aug 1999, arXiv:quant-ph/9908066v1.

\bibitem[KR57]{Kant2}
L.V. Kantorovic and G.~S. Rubinstein.
\newblock On the functional space and certain extremum problems.
\newblock {\em Dokl. Akad. Nauk SSSR (N.S.)}, 115:1058--1061, 1957.

\bibitem[KSV02]{CQinfo}
A.~Yu. Kitaev, A.~H. Shen, and M.~N. Vyalyi.
\newblock {\em Classical and Quantum Computation}, volume~47 of {\em Graduate
  Studies in Mathematics}.
\newblock Amer Mathematical Society, July 2002.

\bibitem[KW10]{KuWe}
Greg Kuperberg and Nik Weaver.
\newblock A new approach to quantum metrics (v1).
\newblock May 2010, arXiv:math.OA/1005.0353.

\bibitem[MS77]{ECC}
F.J. MacWilliams and N.J.A. Sloane.
\newblock {\em The Theory of Error-Correcting Codes}.
\newblock North-Holland Publishing Complany, New York, 1977.

\bibitem[NC00]{NeCh}
Michael~A. Nielsen and Isaac~L. Chuang.
\newblock {\em Quantum Computation and Quantum Information}.
\newblock Cambridge University Press, 2000.

\bibitem[Rai02]{Rains02}
Eric Rains.
\newblock Quantum weight enumerators.
\newblock {\em IEEE}, 44(4):1388--1394, Jul 2002.

\bibitem[RHSS97]{nonadditive}
E.M. Rains, R.H. Hardy, P.W. Shor, and N.J.A. Sloane.
\newblock A nonadditive quantum code.
\newblock Mar 1997, arxiv:quant-ph/9703002.

\bibitem[Rie03]{GHDQMC}
Marc Rieffel.
\newblock Gromov-hausdorff distance for quantum metric spaces.
\newblock Feb 2003, arXiv:math.OA/0011063 v4.

\bibitem[Rie04a]{Rie04}
Marc Rieffel.
\newblock Compact quantum metric spaces.
\newblock {\em Operator algebras, quantization, and noncommutative geometry,
  Contemp. Math.}, 365:315--330, 2004.

\bibitem[Rie04b]{QGHC}
Marc Rieffel.
\newblock Matrix algebras converge to the sphere for quantum gromov-hausdorff
  convergence.
\newblock {\em Mem. Amer. Math. Soc.}, 168(796):67--91, 2004.

\bibitem[Rie10]{LeibQGHC}
Marc Rieffel.
\newblock Leibniz seminorms for "matrix algebras converge to the sphere".
\newblock Jan 2010, arXiv:math.OA/0707.3229.

\bibitem[RR93]{QTAM}
K.~Srinivasa Rao and V.~Rajeswari.
\newblock {\em Quantum Theory of Angular Momentum}.
\newblock Springer-Verlag, 1993.

\bibitem[SL96]{QMac}
P.W. Shor and R.~Laflamme.
\newblock Quantum macwilliams identities.
\newblock Oct 1996, arXiv:quant-ph/9610040.

\bibitem[Tve66]{Tver}
H.~Tverberg.
\newblock A generalization of radon's theorem.
\newblock {\em J. London Math. Soc.}, 41:123--128, 1966.

\end{thebibliography}
\end{document}